\documentclass[11pt]{article}
\usepackage{amsmath,amsfonts, amssymb,graphicx,geometry,authblk,subfigure,algorithm,algorithmic} 
\graphicspath{{Figures/}}

\title{Large scale \textit{ab-initio} simulations of dislocations}

\author[1]{Mauricio Ponga}
\author[2]{K. Bhattacharya}
\author[2]{M. Ortiz}

\affil[1]{\small Department of Mechanical Engineering, University of British Columbia, 2054 - 6250 Applied Science Lane, Vancouver, BC, Canada, V6T 1Z4}
\affil[2]{\small Division of Engineering and Applied Science, California Institute of Technology, 1200 E. California Blvd., Pasadena CA, USA 91125.}%

\begin{document}

\maketitle

\begin{abstract}
We present a novel methodology to compute relaxed dislocations core configurations, and their energies in crystalline metallic materials using large-scale \emph{ab-intio} simulations. The approach is based on MacroDFT, a coarse-grained density functional theory method that accurately computes the electronic structure but with sub-linear scaling resulting in a tremendous reduction in cost.  Due to its implementation in \emph{real-space}, MacroDFT has the ability to harness petascale resources to study materials and alloys through accurate \emph{ab-initio} calculations. Thus, the proposed methodology can be used to investigate dislocation cores and other defects where long range elastic defects play an important role, such as in dislocation cores, grain boundaries and near precipitates in crystalline materials.   We demonstrate the method by computing the relaxed dislocation cores in prismatic dislocation loops and dislocation segments in magnesium (Mg).  We also study the interaction energy with a line of Aluminum (Al) solutes.   Our simulations elucidate the essential coupling between the quantum mechanical aspects of the dislocation core and the long range elastic fields that they generate.  In particular, our quantum mechanical simulations are able to describe the logarithmic divergence of the energy in the far field as is known from classical elastic theory.  In order to reach such scaling, the number of atoms in the simulation cell has to be exceedingly large, and cannot be achieved with the state-of-the-art density functional theory implementations.
\end{abstract}




\section{Introduction}
Dislocations are the main carrier of plasticity in crystalline materials. Their behavior and interactions have strong consequences in the mechanical response of materials. As such, dislocations have been extensively studied since the seminal work of Volterra \cite{Volterra:1907}. Dislocations have long been analyzed by separating their effect into two main contributors, the dislocation core and its long-range field \cite{Hirth}. The former is characterized by strong non-linear  quantum mechanical effects; the latter is a long-range elastic field characterized by smooth atomic displacements that are typically modeled using continuum elasticity theory in either its isotropic or anisotropic versions. 
However, the elastic fields are long-range: the interatomic distances or strains decay inversely with distance $\sim r^{-1}$, where $r$ is the distance from the defect, so that the energy density is logarithmically divergent.  This means that the core and far-field are intimately coupled and the traditional separation is problematic.  
This has many physical implications including the effect of the far-field stresses on the core and the interaction between the dislocation and solute atoms.  Therefore, a full understanding requires  \emph{ab-initio} techniques \cite{Rodney:2017}. However, this is not a trivial task since the slow decay of the far field and the intimate coupling means that a very large number of atoms are needed to properly describe this behavior. Unfortunately, this cannot easily be achieved with existing \emph{ab-initio} methods, since they scale poorly ($\mathcal{O}(N_e^3)$ in traditional approaches and $\mathcal{O}(N_e)$ in newer linear scaling approaches \cite{Parr:1989,gs_cse_03}).

Different multiscale approaches have been proposed to overcome these difficulties. Most of these models consist of patching together heterogeneous models at different scales, from DFT to continuum elasticity to couple the long-range effect and the dislocations core. The coupling between the various models varies from parameter passing to hybrid Hamiltonians \cite{Abraham:1998,Lu:2006,Kermode:2008, Bernstein:2009,Curtin:2010,Zhang:2013}. Since these methods bring together distinct models that embody different physics and differing mathematical formulations, they typically assume separation of scales (as in parameter passing) or additional physics or constraints at the interface (as in hybrid Hamiltonians). Often times, the models need to be adjusted or calibrated to the particular phenomenon under consideration, which ultimately detracts from the predictiveness and fundamental nature of first-principles calculations.  A variation is the interesting work of Woodward and collaborators \cite{Woodward:2005,Woodward:2008} where they used full density functional theory in a core region and a linear response (Green's functions) outside.  Still, the approach separates the domain into multiple regions with two distinct theories.

In this work, we propose a \textit{new approach where the equations of density functional theory are solved on a domain large enough to capture both the core and elastic fields in a completely seamless manner}.  It builds on the recently developed Coarse-Grained Density Functional Theory (CG-DFT) technique developed by Ponga \emph{et. al.} \cite{Ponga:2014,Ponga:2016b} and has it roots in the linear scaling method of Suryanarayana \emph{et al.} \cite{Suryanarayana:2013}.   The latter reformulates the original Kohn-Sham equations \cite{Hohenberg-Kohn,Kohn-Sham} into density matrix form, and then approximates the density matrix using spectral Gauss quadrature rules. This enables the calculation of the electron density and other quantities of interest at any point in space at fixed or $\mathcal{O}(1)$ cost.  Consequently the algorithm has linear scaling.  However, the fact that the method is {\it local} allows the construction of an adaptive numerical approximation that has {\it sub-linear} scaling.  The key idea is to exploit the decay so that we have full resolution near the core where the details are important and only sample the electronic fields far away where the atomistic displacements decay in a smooth fashion.  However, this requires care: even though the atomistic displacements decay smoothly enabling a quasi-continuum approximation \cite{TadmorQC,KnapQC}, the electronic fields (like the electron density and Hartree potential) oscillate in an (almost) periodic manner on the scale of the lattice.  Our key observation is that the electronic fields of interests can be written as a sum of two parts, a {\it predictor} which captures the periodicity in the far-field and a {\it corrector} that is complex at the core but decays away from it.  We compute the former inexpensively via unit cell calculations at representative points, while we compute the latter on a gradually coarsening set of {\it electronic sampling points} by exploiting the local nature of our formulation.  Together, this gives sub-linear scaling with the number of electrons allowing the multiscale simulations of hundreds of thousands of atoms with DFT. The technique has been implemented in the \textit{MacroDFT} code. We will refer to the CG-DFT technique as well as the code as MacroDFT.

The key difference between our prior work \cite{Ponga:2016b} is the treatment of the predictor field.  In our previous work \cite{Ponga:2016b}, the predictor was taken to coincide with the periodic solution of the perfect stress-free crystal.  While this choice is effective for defects that decay quickly, i.e., $r^{-2}$ or $r^{-3}$, it is not for dislocations where the decay is much slower, i.e., $r^{-1}$.    The slower decay requires us to update the predictor non-uniformly depending on the far-field deformation, and this adds significant complexity to the method.  This is accomplished in this work.

We use MacroDFT to compute the formation energy of prismatic dislocation loops and basal dislocations in magnesium (Mg). Dislocations in Mg have been extensively studied using \emph{ab-initio} techniques \cite{Tsuru:2015,Shin:2011,Ghazisaeidi:2014}, molecular statics \cite{Nogaret:2010} or combination of them \cite{Yasi_2009}. However, often times these simulations are carried out using incompatible boundary conditions such as periodic boundary conditions or free surfaces to patch the interface with continuum approaches, that ultimately affects the energy of the dislocation core, and possible its configuration \cite{Rodney:2017}. These conditions do not reflect the true nature of a dislocation core, and can influence the structure of the dislocation core \cite{Cai:2003,Li:2012,Romaner:2014} as in the case of body-centered materials. Due to these artifacts, no simulation technique using \emph{ab-initio}technique has been able to prove the theoretical logarithmic divergence predicted by the elasticity theory. In this work, we endeavor to address some of the limitations from other \emph{ab-initio} techniques and study dislocation cores using compatible boundary conditions, and to investigate the scaling of the energy for large simulation cells. Prismatic dislocations loops are made by removing a certain amount of atoms from the bulk material in the basal plane, leading to a Burgers vector ($b$) in the $c-$component of the hexagonal closed-packed (hcp) structure of Mg. On the other hand, basal and prismatic screw dislocations are characterized by a much shorter Burgers vector, and their far field is characterized by the elastic decay. Thus, we generate a single dislocation in an infinite crystal, we measure the evolution of the strain energy as a function of the distance from the dislocation core. Our simulations show that the strain energy diverges logarithmically with the ratio $r/b$, in agreement with the elasticity theory of dislocations \cite{Hirth}. We then compute the interaction energy of the dislocation core with a row of solute atom, and show the transition barriers as dislocation moves through it.

The manuscript is organized as follows. We start in Section \ref{Methodology} by introducing the methodology in terms of a density matrix formulation and its spectral representation in Riemann-Stieljes integrals. We then introduce the coarse-grained representation. Next, the computational set-up and verification of the method is discussed in Section \ref{SimulationsSetUp}. The results for prismatic loops and screw dislocations are discussed in Section \ref{Results}, and the parallel performance of the method is demonstrated in \ref{parallel}.  Finally, we summarize the work with main outcomes in Section \ref{Conclusions}.

\section{Methodology} \label{Methodology}

\subsection{Density Functional Theory} \label{sec:dft}
Consider a system of $M$ atoms with $N_e$ electrons and let $\mathbf{R} = \lbrace \mathbf{R}_1, \mathbf{R}_2, \ldots, \mathbf{R}_M \rbrace$ denote the position of nuclei with charges $\lbrace Z_1, Z_2, \ldots, Z_M \rbrace$, respectively. The corresponding energy of the system according to the Local Density Approximation of Kohn-Sham DFT\footnote{We ignore spins in this presentation though they are easily incorporated.} is \cite{Parr:1989,Finnis2003}
\begin{equation} \label{eq:dft}
\begin{split}
{\mathcal E} &=   \int_{\mathbb{R}^3} \left( \sum_{n=1}^{N_e/2} |\nabla \psi_n|^2  + e_{xc} (\rho) \right) dx 
+ \frac{1}{2} \int_{\mathbb{R}^3}    \int_{\mathbb{R}^3} \frac{\rho(\mathbf{x})\rho(\mathbf{x}')}{|\mathbf{x}-\mathbf{x}'|}    \, dx \, dx' \\
& \quad \quad
+  \int_{\mathbb{R}^3}    \rho(\mathbf{x})    \left(    \displaystyle\sum_{J=1}^{M} \frac{Z_J}{|\mathbf{x}-\mathbf{R}_j|}    \right)     \, dx
+    \frac{1}{2}  \displaystyle\sum_{I=1}^{M}   \displaystyle\sum_{\substack{ J=1 \\ I \neq J}}^{M}   \frac{Z_I Z_J}{|\mathbf{R}_I-\mathbf{R}_J|},
\end{split}
\end{equation}
where $\psi_n$ are the electronic orbitals, $\rho({\mathbf x}) = 2\displaystyle\sum_n^{Ne/2} |\psi_n({\mathbf x})|^2$ is the charge density and the exchange-correlation is given by
\begin{equation}
 e_{xc} (\rho) =  -\frac{3}{4} \left( \frac{3}{\pi}\right)^{1/3} \rho^{4/3} + \rho \varepsilon_c(\rho),
\end{equation}
with $\varepsilon_c$ to be as proposed by Perdew and Wang \cite{Perdew1992} fitted to accurate Monte Carlo simulations carried out by Ceperley \emph{et. al.} \cite{Ceperley1980}.  The expensive Coulombic double sums are sidestepped by introducing the electrostatic potential $\phi$ as the solution to the Poisson equation \cite{IsmailBeigi20001,Suryanarayana2010}.

We minimize this functional over all orbitals subject to orthonormality and this leads to the nonlinear eigenvalue problem
\begin{equation} \label{Eqn:EigValue:Zero}
    \mathcal{H} \psi_{n} = \lambda_{n} \psi_{n}
    \qquad n = 1, 2, \ldots N_e/2 ,
\end{equation}
with $\lambda_n$ the ordered eigenvalues of the Hamiltonian  
\begin{equation} \label{Eqn:Hamiltonian}
    \mathcal{H} = -\frac{1}{2}\nabla^2 + u + \phi,
\end{equation}
with $\phi$ the Coulomb potential obtained by solving the Poisson's equation
\begin{equation} \label{eq:poisson}    
 -  \frac{1}{4\pi} \nabla^{2}\phi(\mathbf{x},\mathbf{R}) =    \rho(\mathbf{x}) + b(\mathbf{x},\mathbf{R}),
\end{equation}
where $b(\mathbf{x},\mathbf{R}) = \sum_{J=1}^{M} b_{J}(\mathbf{x},\mathbf{R}_J)$ denotes the total charge density of the nuclei, with $b_{J}(\mathbf{x},\mathbf{R}_J)$ representing the regularized charge density of the $J^{\text{th}}$ nucleus, and 
\begin{equation} \label{eq:xc}
u ({\mathbf x}) = {\partial e_{xc} \over \partial \rho} (\rho ({\mathbf x} )).
\end{equation}
This nonlinear eigenvalue problem is solved using fixed point iteration in the self-consistent formulation (SCF).  This leads to an expensive $\mathcal{O}(N_e^3)$ computational procedure that critically restricts the size of the systems to $100-1,000$ electrons in practice. Recent real-space implementations using efficient finite element basis and finite differences can increase this number up to roughly 10,000 electrons \cite{Motamarri:2013,Pratapa:2016,Motamarri:2019}. 

With the goal of obtaining a linear scaling algorithm, we reformulate the problem above using the the density matrix formulation following Anantharaman and Canc\`es \cite{ac_aihp_09} and Wang {\it et al.} \cite{wang}.   The problem of minimizing (\ref{eq:dft}) is equivalent at zero temperature to the variational problem
\begin{equation}
\begin{split}
{\mathcal E} = 
\max_\phi \max_u \min_\gamma \left( \text{Tr } \mathcal{H} \gamma +   \frac{1}{8 \pi}
    \int_{\mathbb{R}^3} (|\nabla\phi|^2 + (\rho+b)\phi + E^*_{xc} (u) )   \, dx  \right. \\
\vspace{2in}  \left. \phantom{ \int_{\mathbb{R}^3}}  -  2 k_B \theta \left( \text{Tr } \gamma \log \gamma  + \text{Tr } (1-\gamma) \log (1-\gamma) \right) \right),
\end{split}
\end{equation} 
where the maximum is taken over $\{ \gamma \in \chi: |\nabla|\gamma|\nabla \in \chi, 0 \le \gamma \le 1, \text{Tr} \ \gamma = N \}$, $\chi$ is the set of all trace class operators on $L^2 ({\mathbb R}^3)$, 
and $E^*_{xc}$ is the Legendre transform of $E_{xc}$.  One can see this formally by setting $\gamma = \sum_{n=1}^{N_e} |\psi_n \rangle \langle \psi_n |$, using the electrostatic potential to rewrite the Coulomb interactions, using a duality transform for the exchange correlation and exchanging the order of extremization.  These steps are explained in detail and justified in \cite{wang}.
It is possible to show that the solution to variational problem is given by the density matrix
\begin{equation} \label{eq:dm}
\gamma = g(\mathcal H; \lambda_f) \quad \text{where} \quad g(\lambda,\lambda_f)
    =
    \frac{1}{1+\exp(\frac{\lambda-\lambda_f}{k_B \theta})},
\end{equation}
and $\lambda_f$ is the Fermi level corresponding to the constraint $\text{Tr} \ \gamma = N $, the electrostatic potential $\phi$ that satisfies the Poisson's equation (\ref{eq:poisson}), the exchange correlation potential $u$ that satisfies (\ref{eq:xc}) and the charge density $\rho( {\mathbf x}) = \gamma({\mathbf x} ,{\mathbf x} )$.

This density matrix formulation is the basis of linear scaling algorithms ($\mathcal{O}(N_e)$).  Typically, the density matrix (\ref{eq:dm}) is expanded using polynomials and the expansion truncated assuming bandedness of the Hamiltonian and the decay of off-diagonal components.  

We follow a different approach, the linear scaling spectral Gauss quadratures (LSSGQ) following Suryanarayana \emph{et al.} \cite{Suryanarayana:2013}.  The basic idea is to appeal to the spectral representation of the operator $\gamma$ and to approximate the spectral integrals using Gauss quadratures rules. For any function $\eta \in L^2 ({\mathbb R}^3)$, 
\begin{equation} \label{eq:9}
\langle \eta | \gamma | \eta \rangle = \int_{\sigma}  g( \lambda; \lambda_F)  d \mu_{\eta \eta} (\lambda) \approx \sum_{k=1}^K g( \lambda_k^\eta ; \lambda_F) w_k^\eta,
\end{equation}
where $\sigma$ is the spectrum of ${\mathcal H}$, and $\mu_{\eta \eta}$ is the spectral measure of ${\mathcal H}$ contracted with $\eta$.   We treat the integral as a Riemann-Stieljes integral and approximate it with quadratures rules with quadrature nodes $\lambda_k^\eta$ and quadrature weights $w_k^\eta$ to obtain the approximation shown.

We use a sixth order finite difference approximation with $N_f$ grid points corresponding to a uniform grid spacing $h$, and choose an orthonormal basis $\{ \eta_p \}$ corresponding to this finite difference basis to represent the Hamiltonian and other quantities.  Then, it is possible to show that the ground state energy is given by \cite{Ponga:2016b,Suryanarayana:2013}
\begin{equation}  \label{eq:tot}
\begin{split}
{\mathcal E} &= 2 \sum_{p=1}^{N_{\rm f}}    \sum_{k=1}^{K}   w_{k}^{\eta_p} \lambda_{k}^{\eta_p}  g(\lambda_k^{\eta_p},\lambda_f) \
	+  h^3 \sum_{p=1}^{N_{\rm f}} (b_p - \rho_p) \varphi_p 
	+  h^3 \sum_{p=1}^{N_{\rm f}} (E_{xc}(\rho_p) - u_p \rho_p)\\
	& \quad - 2 k_B \theta \sum_{p=1}^{N_{\rm f}} \sum_{k=1}^{K}
	w_{k}^{\eta_p} \left( g(\lambda_k^{\eta_p},\lambda_f) \log g(\lambda_k^{\eta_p},\lambda_f)
        + (1-g(\lambda_k^{\eta_p},\lambda_f)) \log (1-g(\lambda_k^{\eta_p},\lambda_f)) \right),
\end{split}
 \end{equation}
with
\begin{eqnarray} \label{eq:Fermi:Rho}
    N_e
    & \approx &
    2 \sum_{p=1}^{N_{\rm f}}
    \sum_{k=1}^{K} w_{k}^{\eta_p} g(\lambda_k^{\eta_p},\lambda_f), \
    \label{Eqn:LSSGQ:Fermi_level:GQ} \\
    \rho_{p}
    & \approx &
    2 h^{-3} \sum_{k=1}^{K} w_k^{\eta_p} g(\lambda_k^{\eta_p},\lambda_f).
    \label{Eqn:LSSGQ:Electron_density:GQ}
\end{eqnarray}

It then remains to determine the quadrature points $\lambda_k^{\eta_p}$ and weights $w_k^{\eta_p}$.  We do so using Lanczos iteration \cite{Golub:2009} (see \cite{Suryanarayana:2013,Ponga:2016b} for details). The {\it key observation} is that since the finite difference basis has finite support, we can compute the the quantities $\lambda_k^p, w_k^p$ with $O(1)$ effort at each nodal point $p$.  So, the method automatically has linear scaling.  Further, this observation means that the evaluation of all the electronic quantities of interest can be done \emph{locally} (once the Fermi level is known), and this enables the coarse-graining approach described below.

Once we compute the electronic fields, we can compute the forces on the atomic nuclei (the variation of the total energy with respect to the nuclear{ positions ${\mathbf R}$) easily using the Hellman-Feynman theorem.

\subsection{Coarse-grained extension} \label{sec:cg}

\begin{figure}[t]
\centering
	\includegraphics[width=0.8\textwidth]{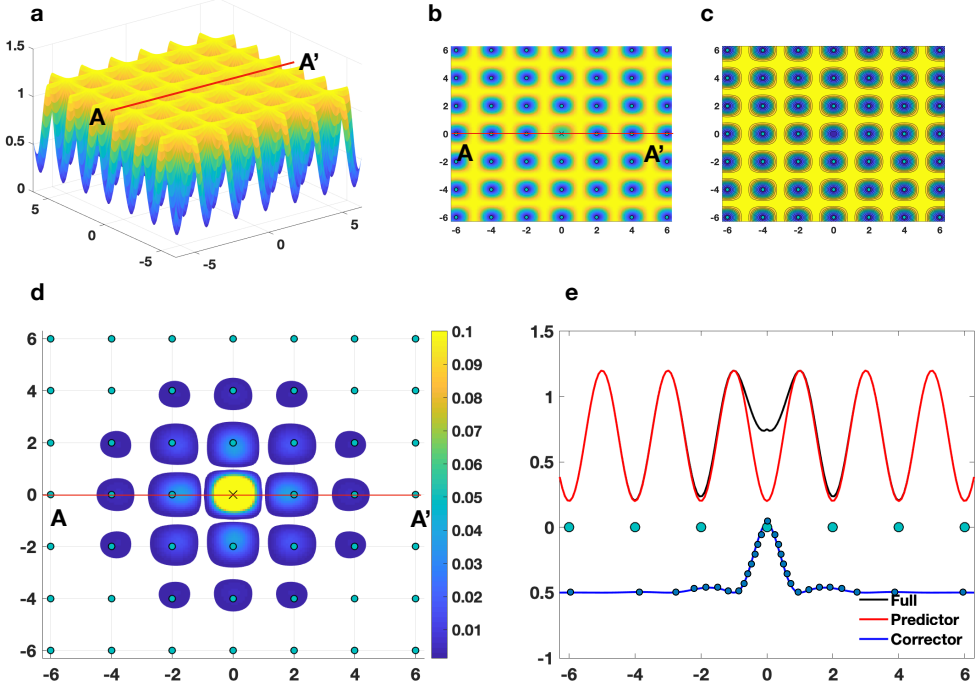}	
\caption{Schematic of an electronic field and its coarse grained representation in MacroDFT. a) Electronic fields around a vacancy defect in a crystalline material. The surface shows the evolution of the field in a 2D plane of the crystalline structure. b) A two-dimensional contour plot of the electronic field ($\rho$). We can clearly see the effect of the vacancy (illustrated with an 'x') in the field. c) Same electronic field for the pristine crystal without defects ($\rho^0$). d) Illustration of the corrector field ($\rho^c = \rho - \rho^0$) for the vacancy example in a two-dimensional slice. Only values above a cutoff has been plotted to ease the view. e) Evolution of the electronic field across the line A-A'. It is evident that the corrector field has large fluctuations near the defect but changes smoothly far away from it. The corrector has been shifted down to make the visualization better. \label{fig:cg}}
\end{figure}

Our goal is a sublinear scaling algorithm that will enable computation of large domains as necessary for dislocations.   We observe that the interatomic distances (or strain) decay at the rate of $r^{-1}$ as we go away from the dislocation core, and the corresponding electronic fields (electron density, electrostatic potential) decay to their periodic behavior at the same rate.  So we introduce a computational basis that makes it possible to describe all details close to the core where it is necessary, but sample it far away where it is not.  Importantly, the approach is seamless and adaptive. Given the different behavior, the coarse graining of the atomic positions and electronic fields are done differently.

\subsubsection*{Coarse grained representation of atoms}
 
The atomistic distances may vary significantly close to the defect, but decay in a polynomial manner.  So, we introduce a subset $\mathcal{P}_{\rm a}$ of atoms, called \emph{representative atoms} (RepAtoms).  This subset is dense, i.e., it contains every atom close to the defect but gradually coarsen away from it.  We track the positions of these atoms and infer the position of other (non-representative atoms) from the position of the representative atoms by introducing {\it linear} finite elements shape functions based on a mesh $\mathcal{T}_{\rm a}$ with nodes $\mathcal{P}_{\rm a}$ as in the atomistic quasi-continuum formulations \cite{Ponga:2016b,TadmorQC,KnapQC,Ariza:2012,Ponga:2015,Ponga:2016,Ponga:2017}. Note that this mesh is Lagrangian. Details of the atomic mesh and visual representation of them are provided in Section \ref{screw-section}.

\subsubsection*{Coarse grained representation of electronic fields}

The electronic fields are complex close to the defect but decay to a periodic field away from it, since the far-field is almost periodic with a period related to the unit cell.  So a smooth interpolation is not appropriate.  We get around it representing the electronic fields as a sum of a {\it predictor} and a {\it corrector}.   The predictor is a slowly varying almost periodic function away from the defect and captures the periodicity of the electronic field in the far field.  The difference between the actual electronic fields and the predictor, therefore smoothly decays away from defect. This is represented by the \emph{corrector.}  Since the corrector decays smoothly, we can use a smooth interpolation away from the defect. 

This is shown schematically in Figure \ref{fig:cg} with a vacancy as the defect. Figure \ref{fig:cg}(a) shows the evolution of an electronic field around the vacancy centered in the middle of the picture. For reference, a line A$-$A' is shown to plot the charge density later on. Next, Figures \ref{fig:cg}(b) and (c) show a contour plot of the charge density for the vacancy and the pristine sample, respectively. We intuitively see that these fields are very similar with the exception of near the defect (marked with an "x"), where large fluctuations are seen. Thus, if we represent the difference between the charge density for the vacancy and the pristine cell, we obtain the contour map shown in Figure \ref{fig:cg}(d). We see that only near the vacancy this field is large, and very quickly decays to zero. This is the \emph{corrector} field, and the pristine solution is called in this case the \emph{predictor}. Putting everything together, we see in Figure \ref{fig:cg}(e) the three fields, i) the full solution with the vacancy ($\rho$) in black; ii) the predictor field ($\rho^0$) in red; and iii) the corrector ($\rho^c$) in blue\footnote{The corrector has been shifted down to make the visualization better.}. Now, it is possible to reconstruct the full solution at any point by using an interpolation scheme provided by the FE mesh. This is schematically shown with the blue markers in the blue line of Figure \ref{fig:cg}(e), where we see a dense sampling near the vacancy, but a coarsening as one moves away from it. This is the principle for coarse-grained description of MacroDFT.

To be precise, we introduce a fine uniform spatial grid $\mathcal{P}_{\rm f}$ and use  representation
\begin{equation} \label{eq:pc}
    \phi_p
    =
    \phi_p^0
    +
    \phi_p^c,
    \qquad
    \rho_p
    =
    \rho_p^0
    +
    \rho_p^c,
    \qquad
    p\in \mathcal{P}_{\rm f} ,
\end{equation}
for the electrostatic field and electron densities at each of these points where $ \phi_p^0, ~\rho_p^0$ are the predictors and $ \phi_p^c, ~\rho_p^c$ are the correctors.  We defer the description of the predictor till the next section.  

To obtain the corrector with sublinear scaling, we compute it only a subset of {\it electronic sampling points} (ESPs) $\mathcal{P}_{\rm c}$.  We then use a finite element mesh $\mathcal{T}_{\rm c}$ over the ESPs and use it to interpolate these fields to the fine mesh:
\begin{equation}\label{phi_approx_scheme}
      \phi_p^c
    =
    \sum_{q\in \mathcal{P}_{\rm c}}
    \gamma(\mathbf{r}_p,\mathbf{r}_q) \phi_q^c ,
    \qquad
    \rho_p^c
    =
    \sum_{q\in \mathcal{P}_{\rm c}}
    \gamma(\mathbf{r}_p,\mathbf{r}_q) \rho_q^c ,
    \qquad
    p\in \mathcal{P}_{\rm f}.
\end{equation}
The ESPs are dense close to the defect, but gradually coarsen away from it.  Note that this mesh is spatial or Eulerian.

Finally, note that we have two meshes, one atomistic and Lagrangian that is used to describe the relaxation of the atoms as the forces and energy are minimized; and an electronic mesh that Eulerian to solve the Kohn-Sham equations in the sample. Both meshes have adaptive resolution, as previously described. They interact through the theory and therefore, it is important to introduce two special functions that map the ESP into the elements of the $\mathcal{T}_{QC}$, and positions to the elements of the fine mesh $\mathcal{T}_{f}$. For more details, the reader is referred to \cite{Ponga:2016b}.

\subsubsection*{Total energy}

\begin{algorithm}[t]
\caption{MacroDFT algorithm}\label{alg}
\begin{algorithmic} \label{Algorithm:1}
   \STATE {\bf initialize.} Initial guess for the positions of RepAtoms $\mathcal{P}_{\rm a}$ and electronic fields at ESPs $\mathcal{P}_{\rm c}$.
\REPEAT
   \STATE Use the atomistic mesh $\mathcal{T}_{\rm a}$ to find position of all atoms,
   \STATE Generate the predictor fields $\phi^0_p, \rho^0_p$ for each $p\in \mathcal{P}_{\rm f}$,
   \REPEAT 
   	\STATE Find the corrector fields $\phi^c_p, \rho^c_p$ using (\ref{phi_approx_scheme}) for each $p\in \mathcal{P}_{\rm f}$, 
	\STATE Find $\{ \lambda_k^{q} \}$ and $\{ w_k^{q} \}$ using Lanczos algorithm for each $q \in \mathcal{P}_{\rm c}$,
	\STATE Find $N_e, \rho_q$  for each $p\in \mathcal{P}_{\rm f}$ using (\ref{Eqn:LSSGQ:Fermi_level:GQ}), (\ref{Eqn:LSSGQ:Electron_density:GQ}),
   \UNTIL{convergence of the electronic fields,}
   \STATE Compute the forces on the RepAtoms $\mathcal{P}_{\rm a}$,
   \STATE Update the position of the RepAtoms $\mathcal{P}_{\rm a}$,
 \UNTIL{convergence of the atomic positions.}
\end{algorithmic}
\end{algorithm}

We compute the atomic positions and electronic fields according to the Algorithm \ref{alg}.  It remains to calculate the total energy (\ref{eq:tot}) of the system.  We do so by sampling using cluster as follows.  Let $\mathcal{C}_q$  be the set or {\it cluster} of nodal points on the fine mesh contained in a ball of radius $r$ around the $q^{\text{th}}$ ESP.  We approximate the total energy (\ref{eq:tot}) as \cite{Ponga:2016b}
\begin{equation} \label{eq:free_cg}
\mathcal{E} \approx \sum_{q \in \mathcal{P}_c } n_q \overline{\mathcal{E}}_q 
\end{equation}
where $n_q$ is the weight of the $q^{\text{th}}$ ESP and it is defined as
\begin{eqnarray}
n_q &=& \begin{cases}
1 & \text{if $q$ belongs to the full resolution zone} \\
\# \mathcal{P}_f \mbox{ associated with $q$th ESP} & \text{otherwise} 
\end{cases}
\end{eqnarray}
and $\overline{\mathcal{F}}_q$ is the average Free-Energy at the $q^{\text{th}}$ ESP computed as
\begin{equation} \label{eq:free_cg_sampling}
\overline{\mathcal{F}}_q = \dfrac{1}{N} \sum_{p \in C_q} \mathcal{F}_p.
\end{equation}

\subsection{Selection of the predictor field} \label{sec:pred}

The \emph{key} for an efficient and accurate coarse-grained implementation is the choice of the predictors.  In our previous work \cite{Ponga:2016b}, the \emph{predictor field} was taken to coincide with the periodic solution of the perfect stress-free crystal. While this choice is effective for defects that decay quickly, i.e., $r^{-2}$ or $r^{-3}$, where $r$ is the distance from the defect, it is not for dislocations where the decay is much slower, i.e., $r^{-1}$.   Therefore, we need a more accurate approximation that reflects the deformation.  

Recall the triangulation $\mathcal{T}_{\rm QC}$.  In each element of this triangulation, the positions of the atoms is obtained from a linear interpolation and thus periodic with a unit cell deformed by the deformation gradient given by the atomistic element.  It is important that the predictor reflect this periodicity.  Therefore, we extract the deformation gradient from each atomistic element, deform the unit cell with this deformation gradient, and compute the electronic field associated with this distorted unit cell.  We then extend this periodic field to the entire element.  Note that this field automatically reflects the local strain.  However, it can be discontinuous at the element boundaries; we remedy this by smoothing with a $L^2 \to H^1$ projection. 

Thus, once the atomic positions are computed with the non-linear conjugate gradient, a new deformation gradient, $\mathbf{F}_e$, is computed per each element. The deformation gradient is then applied to a unit cell, and the electronic fields are computed on it. We then extract the corrector fields, i.e., $\rho^0(\mathbf{x})$ and $\phi^0(\mathbf{x})$, and map them back to the full simulation as schematically shown in Figure \ref{Fig:Predictor}. The calculations in the unit cell are carried out by individual nodes, and their results are stored on the disk to save memory with an MPI-file. While this puts some restrictions to the speed of the code, the time is not comparable with the amount taken to solve the DFT equations, and then it is affordable.

\begin{figure} 	
\centering
	\includegraphics[width=0.5\textwidth]{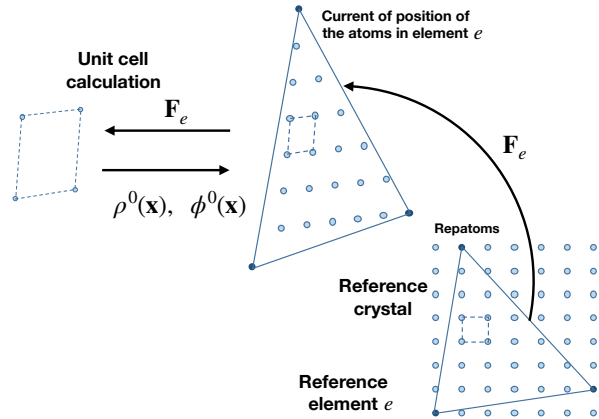}	
	\caption{Schematic showing the evaluation of the predictor fields using Algorithm \ref{Algorithm:1}. The deformation gradient $\mathbf F_e(\mathbf x)$ is extracted from the mesh over the RepAtoms (dark). Non-representative atoms are shown in light. With this information, a periodic calculation using the unit cell is performed and the electronic fields of interest are extracted and mapped forward to the simulation. \label{Fig:Predictor}}
\end{figure}

\section{Computational set-up} \label{SimulationsSetUp}
We now describe the details of the simulations. The details of the meshes used in this work are described in Table \ref{TableCG}. 

\subsection{Verification of the method}

\begin{figure}
\centering
\subfigure[Convergence with respect to quadrature points]{\includegraphics[width=0.45\textwidth]{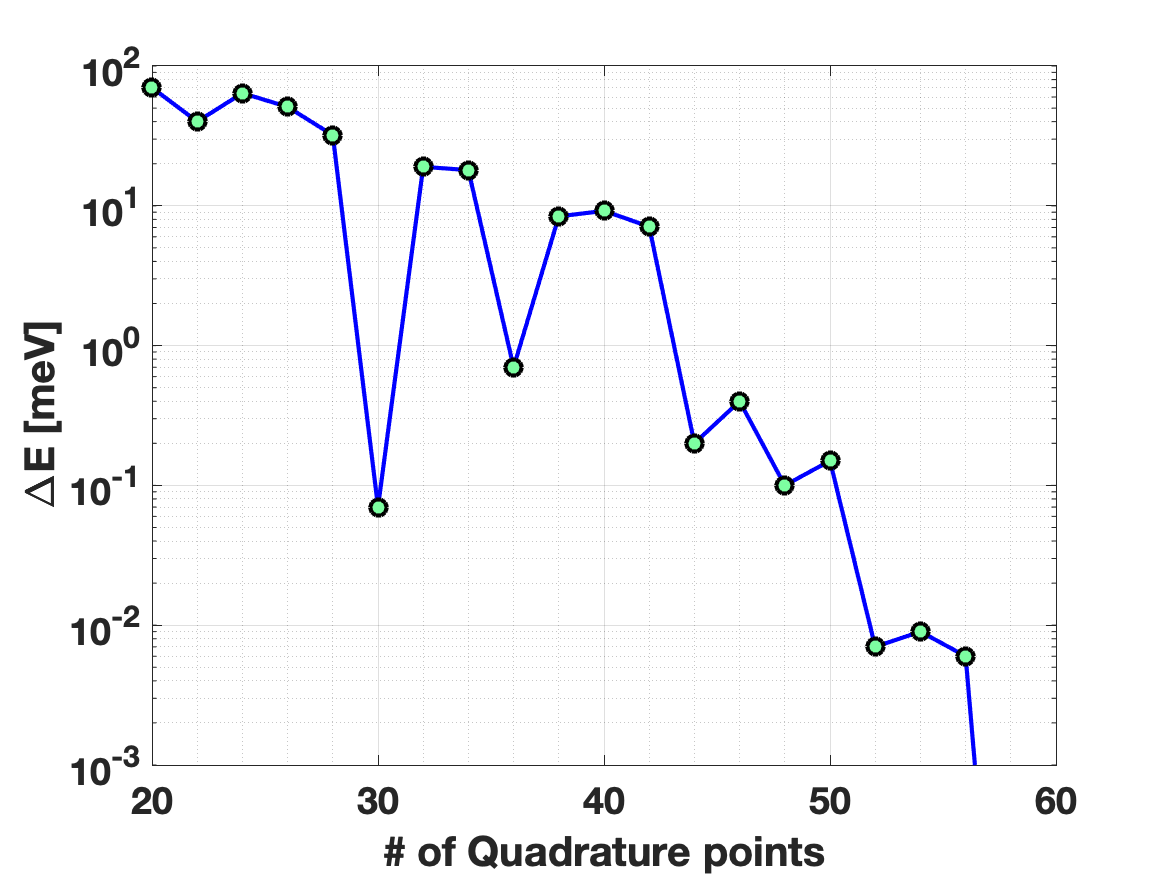}}
\subfigure[Convergence with respect to spatial discretization]{\includegraphics[width=0.45\textwidth]{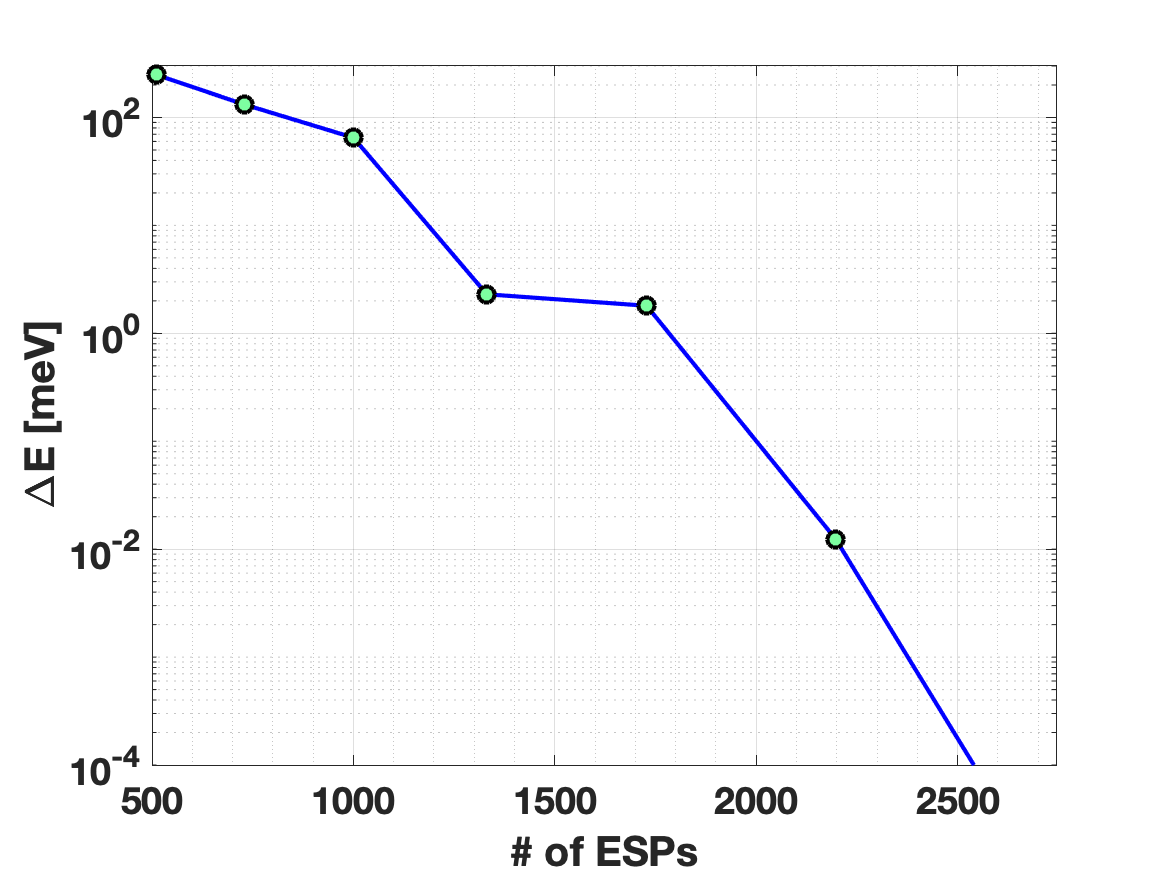}}
\caption{\label{Fig::0} Convergence of the cohesive energy in Mg as a function of the (a) quadrature numbers, and (b) the number of ESPs in the unit cell. The relative energies are compared with respect to a reference simulation done with NQP = 60 and 2744 ESPs.}
\end{figure}

Verification of our method has been provided in our previous publication \cite{Ponga:2016b}, and here we reproduce some of these results for completeness. In our formulation, we first start by investigating two main parameters that condition the accuracy and convergence of the methodology. These two parameters are the number of quadrature points, used to compute Eqs. \ref{eq:tot} and \ref{eq:Fermi:Rho}, and the number of ESP used to discretize the atomic unit cell in full resolution. Figure \ref{Fig::0} shows the convergence with respect to number of quadrature points and spatial discretization for a unit cell of Mg. We see how the difference in the energy is reduced as both quadrature points and number of ESPs is increased. In this work, we retained the spatial resolution we used in our previous work. This is about 2,300 ESPs per unit cell, which ensures a convergence of about 0.01 meV per unit cell. We have also used a sufficiently large number of quadrature points, $K = 40$. With these parameters, is possible to achieve cohesive energy for Mg that are converged to less than a 1 meV. Using a unit cell and full resolution, we have computed cohesive energy values, bulk modulus, and reference volume for Mg that are in good agreement with previous works that have used plane wave implementations. These values are shown in Table \ref{Table:Validation} for completeness. We can see that the values obtained with MacroDFT are in close agreement with ABINIT calculations. The stable stacking fault energy for Mg was also computed with MacroDFT using the described parameter in a simulation cell containing $2\times 2\times 48$ unit cells, i.e., 192 atoms. The values of the stable stacking fault energy was found to be $\sim 27$ mJ$\cdot$m$^{-2}$, in good agreement with previous works \cite{Yasi_2009}. Convergence of our coarse-grained simulations has been done in our previous work \cite{Ponga:2016b}. In addition, MacroDFT has been used to compute the energy of twin boundaries in Mg, showing good agreement with other techniques \cite{Sun:2018}. Here, we used the same level of coarsening, that is, the element size increases in different regions of coarse representation by a factor of two. This ensures that the evaluated energy does not depend of the mesh size. 

\begin{table}
\centering \caption{Comparison of cohesive energy, bulk modulus, and reference volume obtained with MacroDFT and ABINIT using the same local pseudopotential.}
\begin{tabular}{c c c}
\hline \hline	
Property & This work & ABINIT \\
\hline 			
E$_{\text{min}}$ [eV] & -24.61	&  -24.678  \\
B$_0$ [GPa] & 38.74	&  38.4  \\
V$_{0}$ [eV] & 42.317	&  42.351  \\
\hline \hline	
\end{tabular}
\label{Table:Validation}
\end{table}

\subsection{Prismatic dislocation loops in magnesium} \label{PDL-section}

A single crystal Mg was generated using the minimum energy configuration for the bulk crystal at $a = { 3.109} $ \AA ~ {and} $c/a = 1.626$ \cite{Ponga:2016b}. The overall length of the computational domain was $40 a_0 \times 40 a_0 \times 40 c_0 $ containing $N = 256,000$ atoms. A full resolution zone of $8a_0 \times 8a_0 \times 8 c_0$ was provided in the center of the simulation cell. This area provided fully resolved atomic and electronic mesh containing around $\sim 1,126,400$ ESPs. Surrounding the full resolution zone, multiple coarse-grained regions were provided for both the electronic and atomic fields. Two Delaunay triangulations over the ESPs and RepAtoms were performed, and a FE mesh was provided over these triangulations to interpolate atomic positions and corrector fields. From the center of the full resolution area, $N_v$ atoms were removed to generate a cluster of vacancies. The boundary conditions were set up such that the solution decay to the pristine solution at the end of the simulation cell, for both atomic positions and electronic fields. Let $\mathbf{u}(\mathbf{R})$ be the atomic displacement field, then $\mathbf{u}(\mathbf{R}) = \mathbf{0}, ~\forall~ \mathbf{R} \notin \mathcal{P}_{QC}$ and $ \rho_c({\mathbf x}) = 0, ~\phi_c({\mathbf x}) = 0 ~\forall~ {\mathbf x}~ \notin \mathcal{P}_{f}$. This means that we enforced the displacement of the atoms to be zero out of the simulation box and forced the corrector field of the electron density and electrostatic potential to be zero outside the simulation cell too. The selection of these boundary conditions is justified based on the idea that the deviations of the electronic field and atomic displacements decay to zero far away of the defect. 

\begin{table}
\centering \caption{Details of the computational meshes and resolution of dislocation cores simulated in this work. '\# Atoms' is the total number of atoms in the crystal, and '\# RepAtoms' is the total number of representative atoms in the simulation.  \#EPs is the total number of electronic points in the fully resolved crystal, and \#ESPs is the total number of ESPs in the simulation cell. The columns '\# Threads' and '\# MPI processes' refer to the number of threads and MPI processes used during the simulation. }
\begin{tabular}{c c c c c c c}
\hline \hline	
Simulation			& \# Atoms	& \# RepAtoms	& \#ESPs	& \#EPs	& \# Threads & \# MPI processes \\
\hline 			
P.D.L.			&   256,000	&  5,000 	& 1,126,400 & 262,144,000	& 64	& 1 \\
Screw Basal		&   158,000	& 4,000 	& 560,395	& 163,840,000 & 64	&  1 \\
Screw Prismatic	&   158,00	0	& 4,000 	& 560,395	& 163,840,000 & 64	&  1  \\
\hline \hline	
\end{tabular}
\label{TableCG}
\end{table}

\subsection{Screw dislocations in magnesium} \label{screw-section}

Two screw dislocations were generated in a single Mg crystal of dimensions $ a_0 [11\overline{2}0] \times  200 \sqrt{3} a_0 [10\overline{1}0]  \times 200 c_0 [0001] $. A basal $\dfrac{a_0}{3}[11\overline{2}0] \{ 0001\}$ and a prismatic $\dfrac{a_0}{3}[11\overline{2}0] \{ 10\overline{1}0\}$ screw dislocations were generated by displacing the atoms in the simulation cell using the isotropic elastic solution \cite{Hirth}. The dislocations were generated such that the dislocation line was aligned with the $x-$ direction in the sample as shown schematically in Figure \ref{Fig:3}.
\begin{figure} 
\centering
\includegraphics[width=0.75\textwidth]{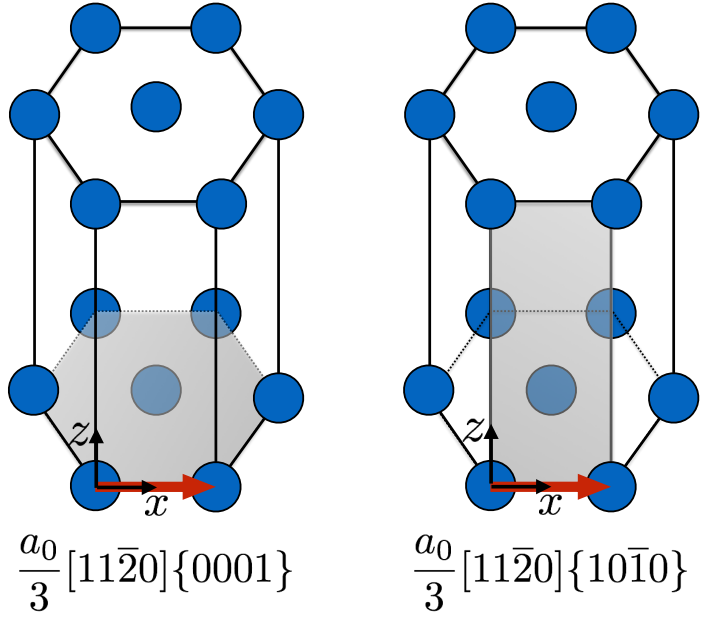}
\caption{Schematic view of the slip systems simulated in this work. The burgers vector of the dislocation is shown in red. The gray shaded are shows the slip plane. Blue spheres denote the position of the atoms in the pristine structure of hcp materials. }
\label{Fig:3}
\end{figure}
A full resolution zone of $a_0 \times 16\sqrt{3} a_0 \times 4 c_0$ surrounding the dislocation core was provided for the $\dfrac{a_0}{3}[11\overline{2}0] \{ 0001\}$. By contrast, a full resolution zone of $a_0 \times 4\sqrt{3} a_0 \times 16 c_0$ surrounding the dislocation core was provided for the $\dfrac{a_0}{3}[11\overline{2}0] \{ 10\overline{1}0\}$. In the full resolution region, the spatial resolution for the electronic degree of freedom is about $2,300$ ESPs per unit cell. The total number of ESPs in the full resolution area is around $281,600$. Surrounding this area, three coarse-grained regions were provided. Each coarse-grained region had a constant element size that increased with the distance to the core. The grand total number of electronic degrees of freedom was around 560,395 ESPs. Over this set of points, a Delaunay triangulation was performed and a FE mesh was generated by using four-nodes tetrahedral elements with linear interpolation functions.

\begin{figure}
\centering
\subfigure[Atoms and mesh for the basal system.]{\includegraphics[width=0.5\textwidth]{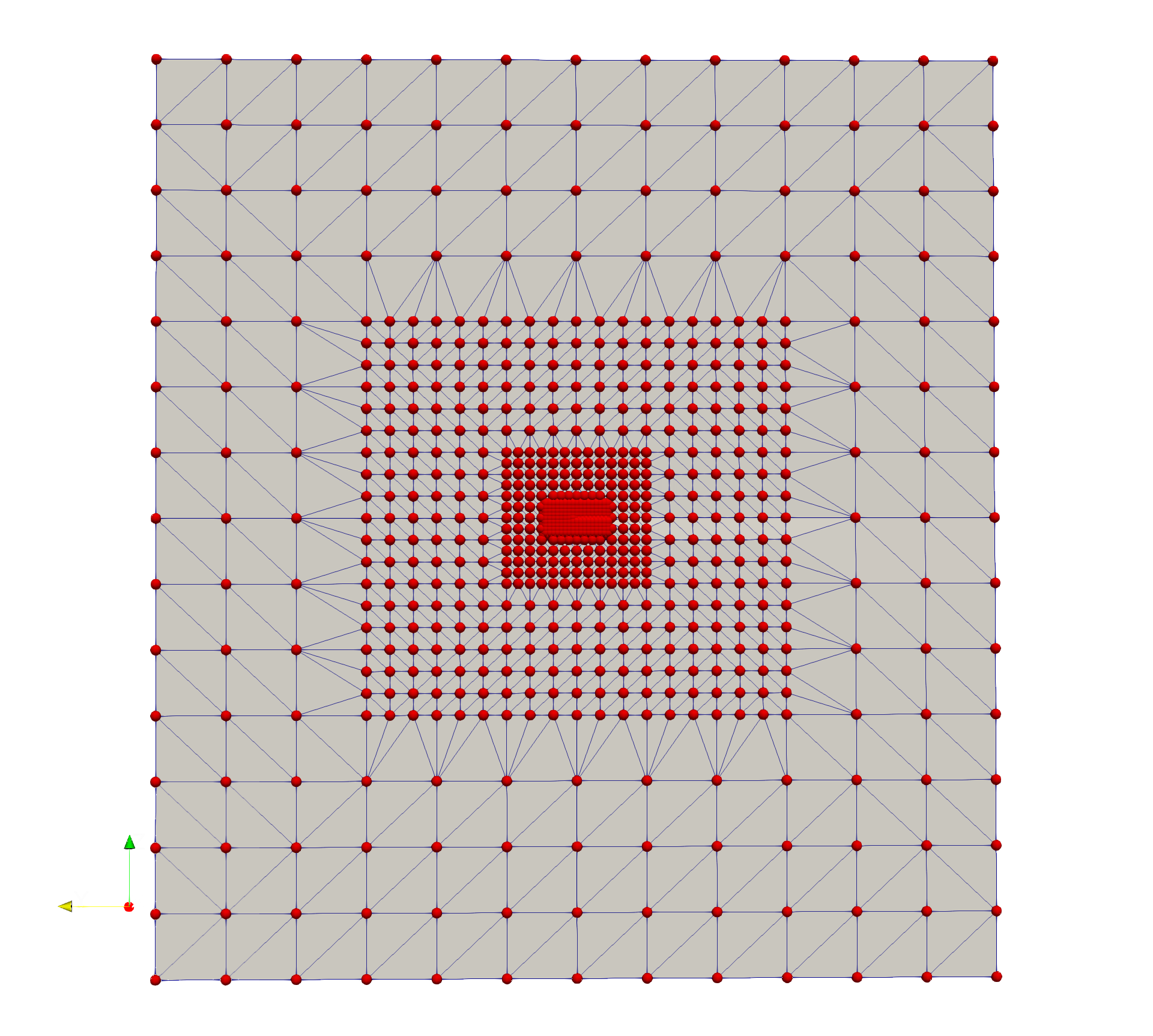}} \\
\subfigure[Full resolution area for the basal system.]{\includegraphics[width=0.45\textwidth]{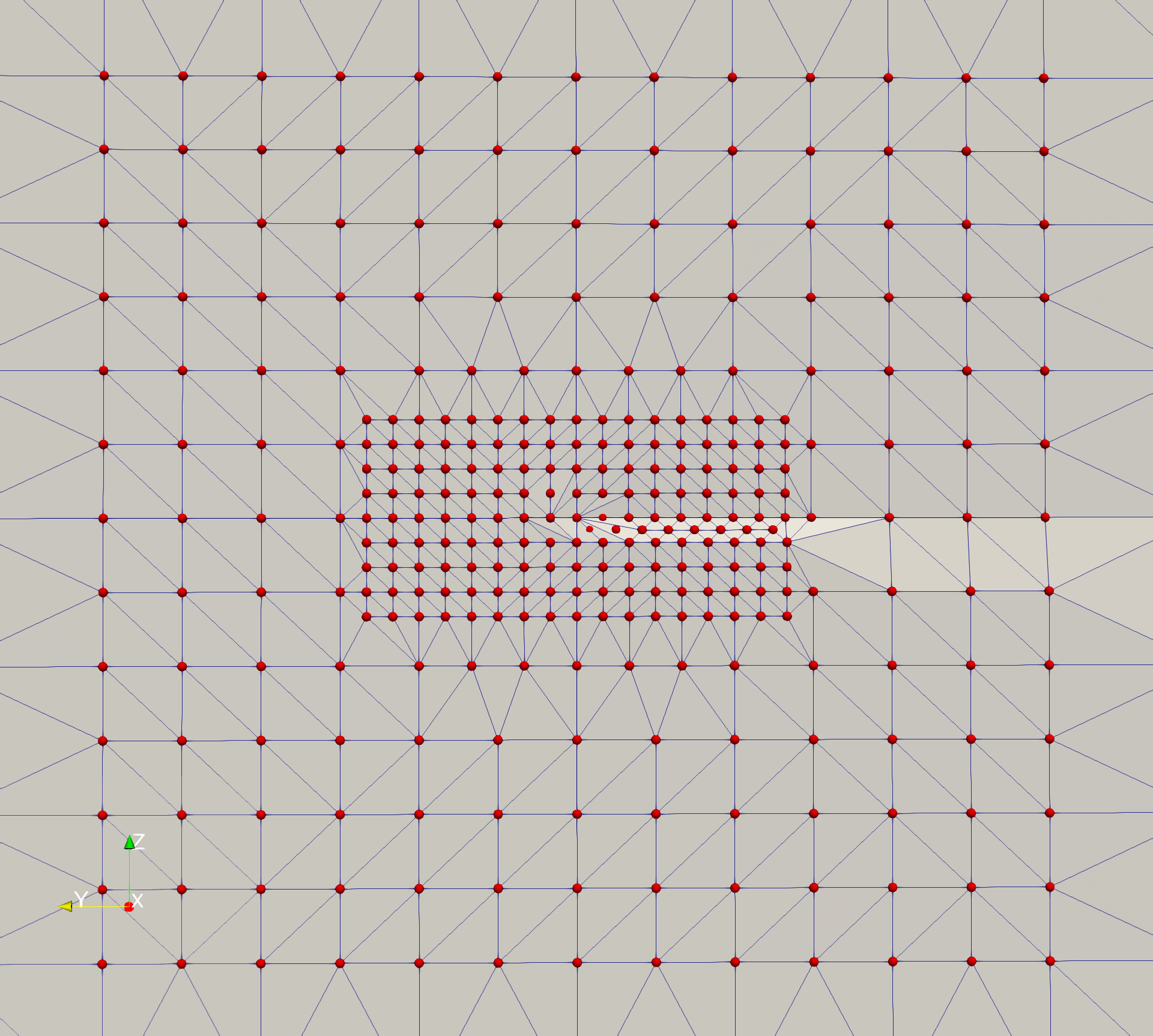}}
\subfigure[Full resolution area for the prismatic system.]{\includegraphics[width=0.45\textwidth]{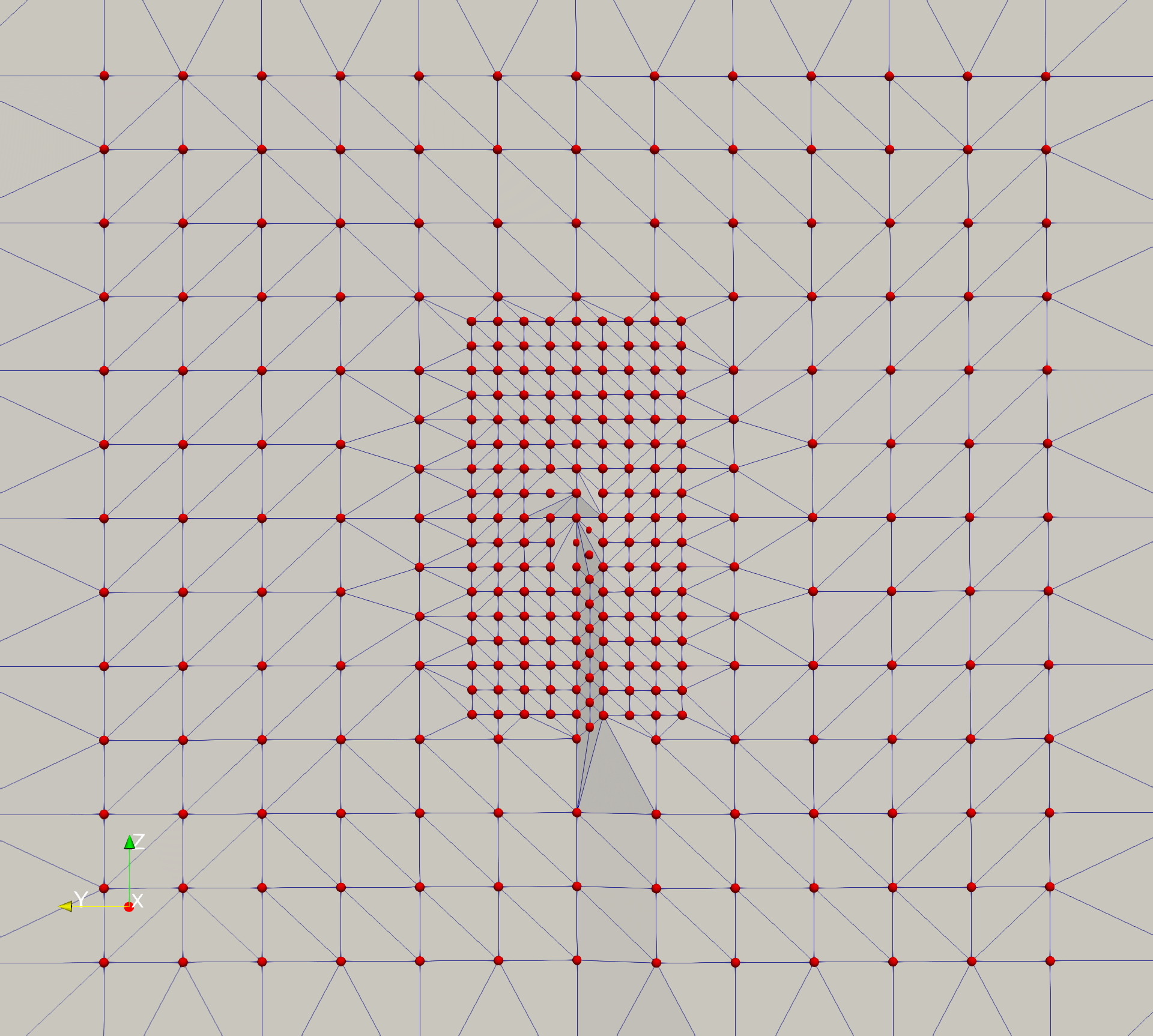}}
\caption{\label{Fig::1} Finite element mesh along with the representative atoms for the $a_0/3[11\overline{2}0]\lbrace10\overline{1}0\rbrace$  screw dislocation simulation. Bottom close-up show the full resolution zone for both slip systems. }
\end{figure}

Similarly to the electronic mesh, an atomic mesh was also provided. The atomic mesh also contained a full resolution area that coincides with the full resolution electronic region. Away from the dislocation core, multiple coarse-grained regions were provided. The perfect atomic positions are used as starting point for this mesh.  Figure \ref{Fig::1}(a) shows spatial distribution of the atoms for a $\dfrac{a_0}{3}[11\overline{2}0] \{ 10\overline{1}0\}$ screw dislocation as well as the FE mesh for a sample with three coarse-grained regions. Figure \ref{Fig::1}(b) and (c) show a close-up of the full resolution zone for the two slip systems considered in this work.

 
  The coarse grained atoms on the slip plane need to have a special capability -similar to the discontinuous FE Galerkin formulation- where the elements are allowed to have a \emph{dual position} in order to interpolate properly the atomic positions in both sides of the slip plane. With the screw dislocation generated, a FE mesh was constructed over the representative atoms using an $\alpha$-Delaunay triangulation. The $\alpha$-Delaunay triangulation was needed because the set of representative atoms with the screw dislocation is not a convex hull and highly distorted elements were generated with a regular Delaunay triangulation. A full resolution simulation of the previously described dislocation will involve 304 million ESPs. This leads to a total acceleration due to the spatial coarse-grained approach of 545 times. 

For all simulations we used a $k_B \theta = 0.8$ which is consistent with other DFT studies of metals and the energy and forces were minimized by using the Polak-Ribi\`ere version of the  non-linear conjugate gradient (NLCG). The boundary conditions were set up such that the atomic displacements outside the computational cell were given by the elastic solution \cite{Hirth}. By doing so, we can compute for all elements outside the computational box the deformation gradient $\mathbf{F}_e$ \textit{a priori}. Thus, for each atom outside the computational domain, the electronic field and atomic displacements are fully determined. This means that $ \rho_c({\mathbf x}) = 0, ~\phi_c({\mathbf x}) =0 ~\forall~ {\mathbf x}~ \notin \mathcal{P}_{f}$. 

\section{Results} \label{Results}

%
%
%

\subsection{Prismatic dislocation loops in magnesium}

The energy of the computational cell is minimized using a NLCG and the formation energy and the binding energy were computed as

\begin{equation}  \label{eq:formation_energy}
E^{f}_{N_v} = E(N-N_v) - E_p(N - N_v) = E(N-N_v) - (N - N_v) E_c,  
\end{equation}
and
\begin{equation}  \label{eq:binding_energy}
E^{b}_{N_v} = N_v E^{f}_{1} -E^{f}_{N_v} ,
\end{equation}
where $E^{f}_{N_v}$ is the formation energy for the cluster with $N_v$ vacancies, $E(N-N_v)$ is the total energy of the system containing $N_v$ vacancies, $E_p(N - N_v) = (N - N_v) E_c$ is the energy in the pristine configuration computed as the number of atoms ($N - N_v$) times the cohesive energy per atom ($E_c$) in Mg. $E^{b}_{N_v}$ is the biding energy for $N_v$ vacancies, and $E^{f}_{1}$ is the formation energy of a single vacancy.

Table \ref{TableComp} shows the values of the formation energy and the binding energy for both the non-collapsed and collapsed configurations. The binding energy for cluster of vacancies lying on the basal plane is quite large and correspond to a very stable defect. The binding energy computed using Eq. \ref{eq:binding_energy} measures the difference between the energy in a cluster of vacancies and the energy of the same number of individual vacancies. Therefore, such a large binding energy indicates that vacancies will always try to regroup in clusters instead of being randomly disperse in the sample, at least at low temperatures where entropic effects are not important. Figure \ref{non-collapsed-7-vac} shows the electron density through two planes at $y = 0$ and $z=0$, respectively for a sample containing 7 vacancies in the non-collapsed configuration. 
 
\begin{figure} 	
\centering
\subfigure[]{\includegraphics[width=0.3\textwidth]{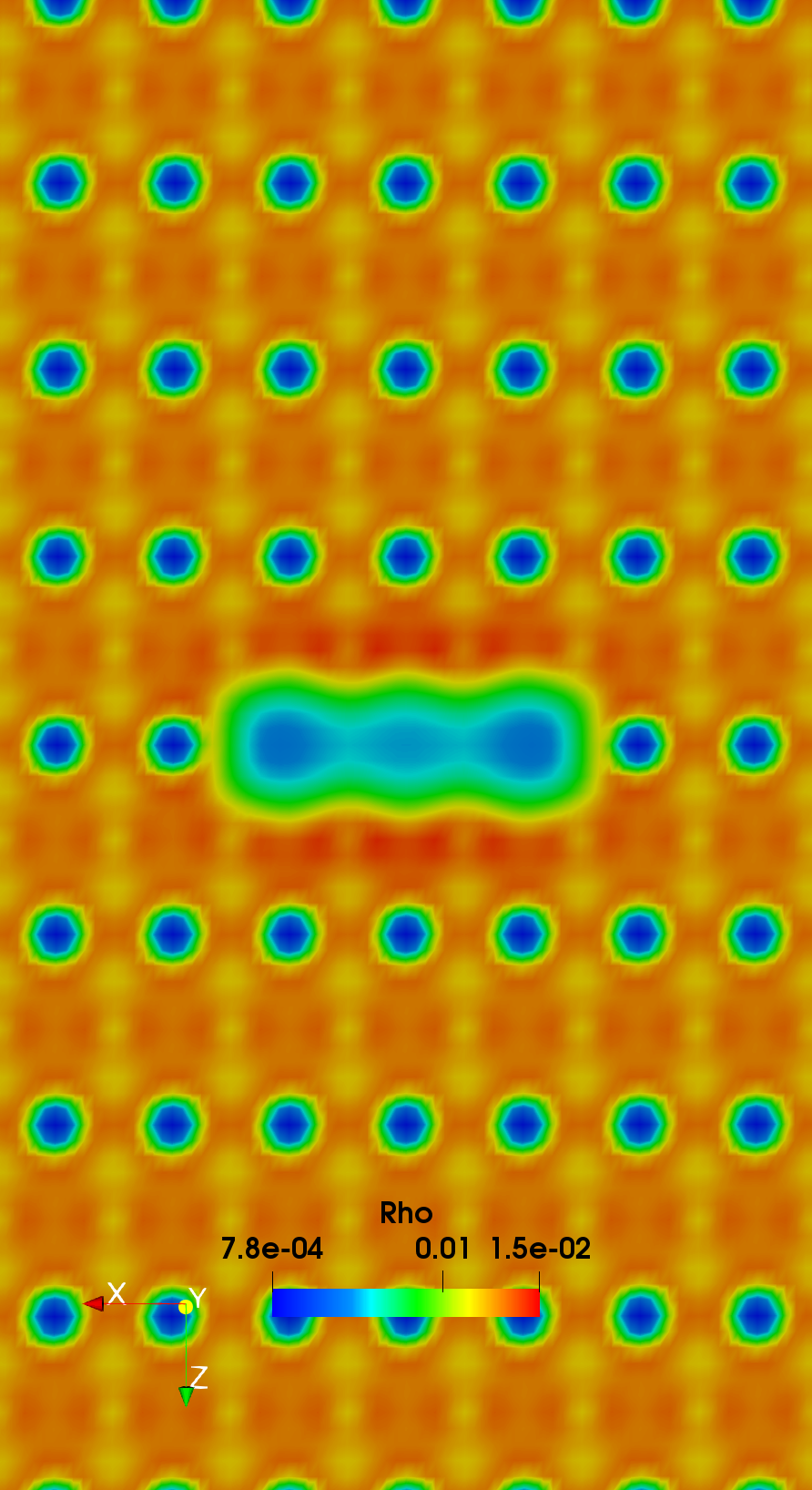}} 
\subfigure[]{\includegraphics[width=0.3\textwidth]{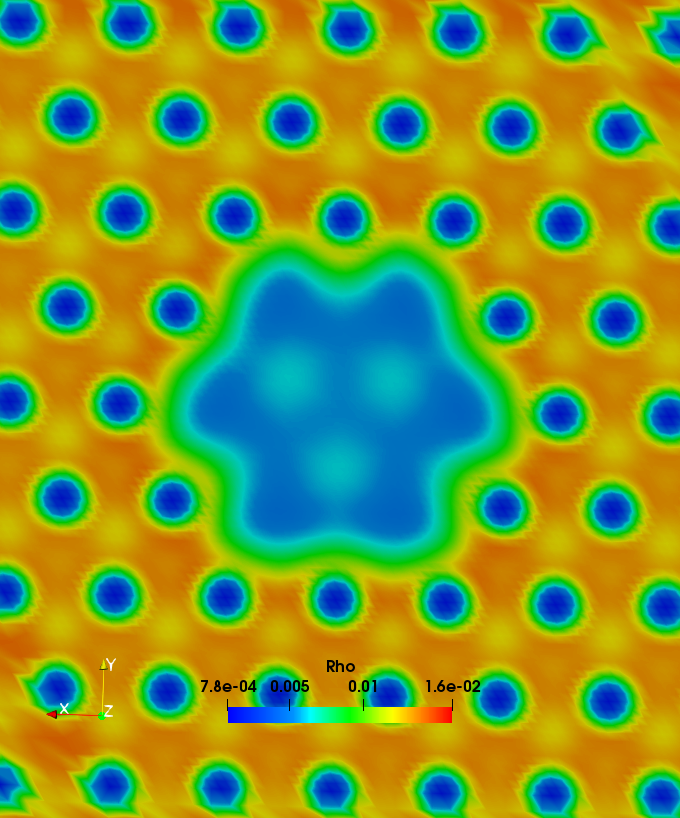}}
\caption{Electron density through two planes at (a) $([11\overline{2}0])$ and (b) $[0001]$ crystallographic directions for a sample containing 7 vacancies in the non-collapsed configuration.  \label{non-collapsed-7-vac} }
\end{figure}

It is also interesting to analyze the binding energies of the same clusters collapsed, in order to understand the formation of prismatic dislocation loops in Mg. For instance, for the sample containing 7 vacancies, the collapsed configuration was always unstable and the atoms came back to the non-collapsed configuration shown in Figure \ref{non-collapsed-7-vac}. On the other hand, when the number of vacancies was increased to $N_v = 19$ and $N_v = 37$, the binding energy experienced a change, as shown in Table \ref{TableComp}. The binding energy for $N_v = 19$ was around 2.99 eV, and for $N_v = 27$ was around 10.13 eV. This highly non-linear demeanor of the binding energy indicates a preference towards large prismatic dislocation loops and therefore, sets a minimum size of PDL that can be observed in Mg. Although configurations with larger number of vacancies are not feasible with MacroDFT, there is a clear trend in the binding energy which favors large PDLs, as observed in experiments \cite{LoopsMg,LoopsMg2,Loops:4} of Mg and nanoporous Mg alloys.

\begin{table}
\centering \caption{Binding energy for PDLs of different sizes after relaxation. In all cases, the number of atoms in the simulation is approximately 256,000. }
\begin{tabular}{c c c c}
\hline \hline	
  \multicolumn{1}{c}{} & \multicolumn{3}{c}{Cluster Size [Atoms]}\\
\hline 	& 7		& 19	& 37   \\
Binding energy (non-collapsed) [eV]			& 1.25 & 4.39 & 9.36 \\
Binding energy (Collapsed) [eV]			& ---  & 2.99 & 10.13  \\
\hline \hline	
\end{tabular}
\label{TableComp}
\end{table}

\begin{figure} 	
\centering
\subfigure[]{ \includegraphics[width=0.33\textwidth]{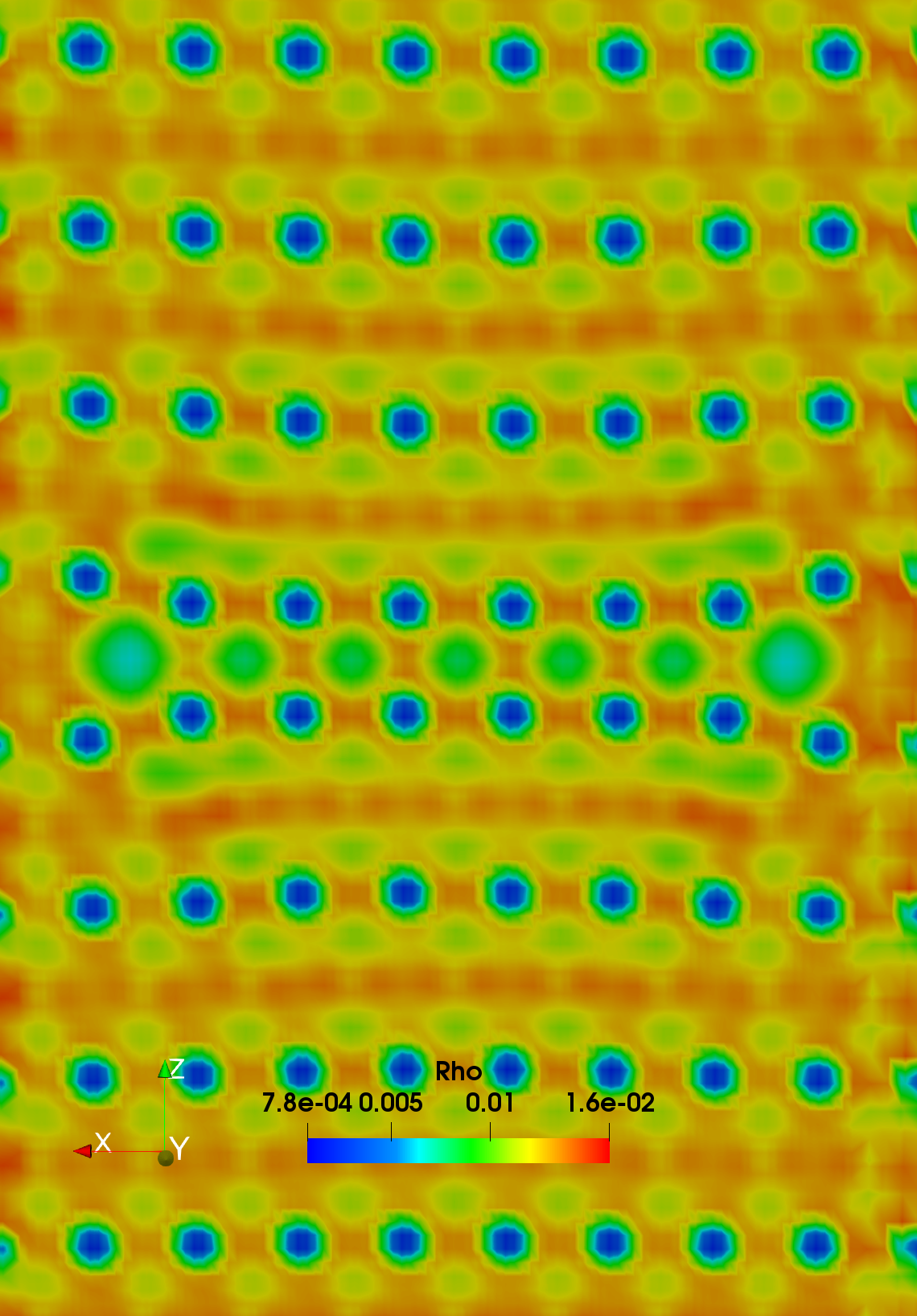}}
\subfigure[]{ \includegraphics[width=0.33\textwidth]{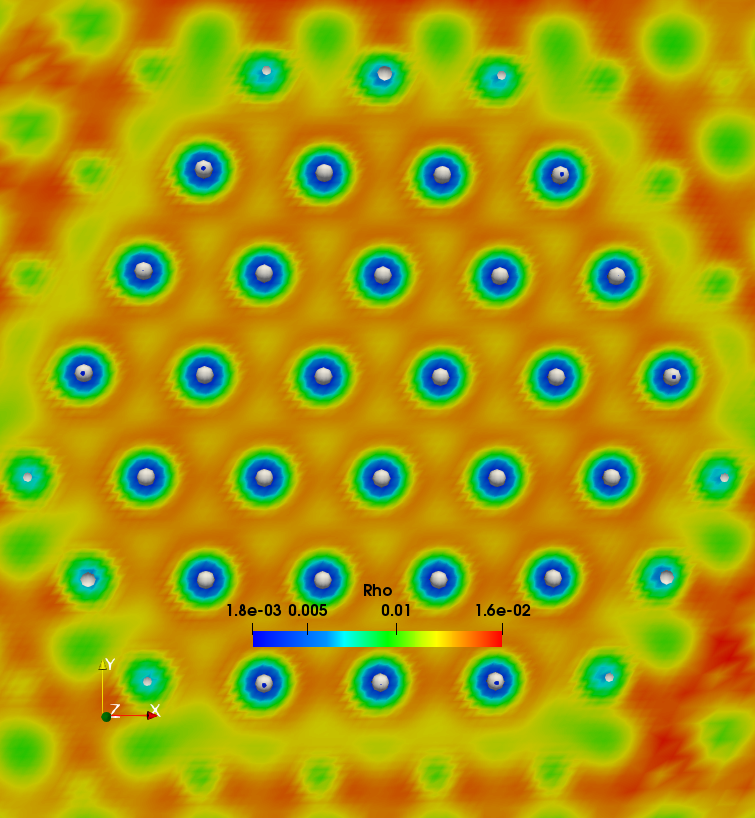}}
\caption{PDLs after collapse and energy relaxation. a) Electron density near the defect, b) top view of the electron density for the collapsed loop.  Atoms that generate the PDL are illustrated with white spheres. \label{pdls}}
\end{figure}

\begin{figure} 	
\centering
	\includegraphics[width=0.75\textwidth]{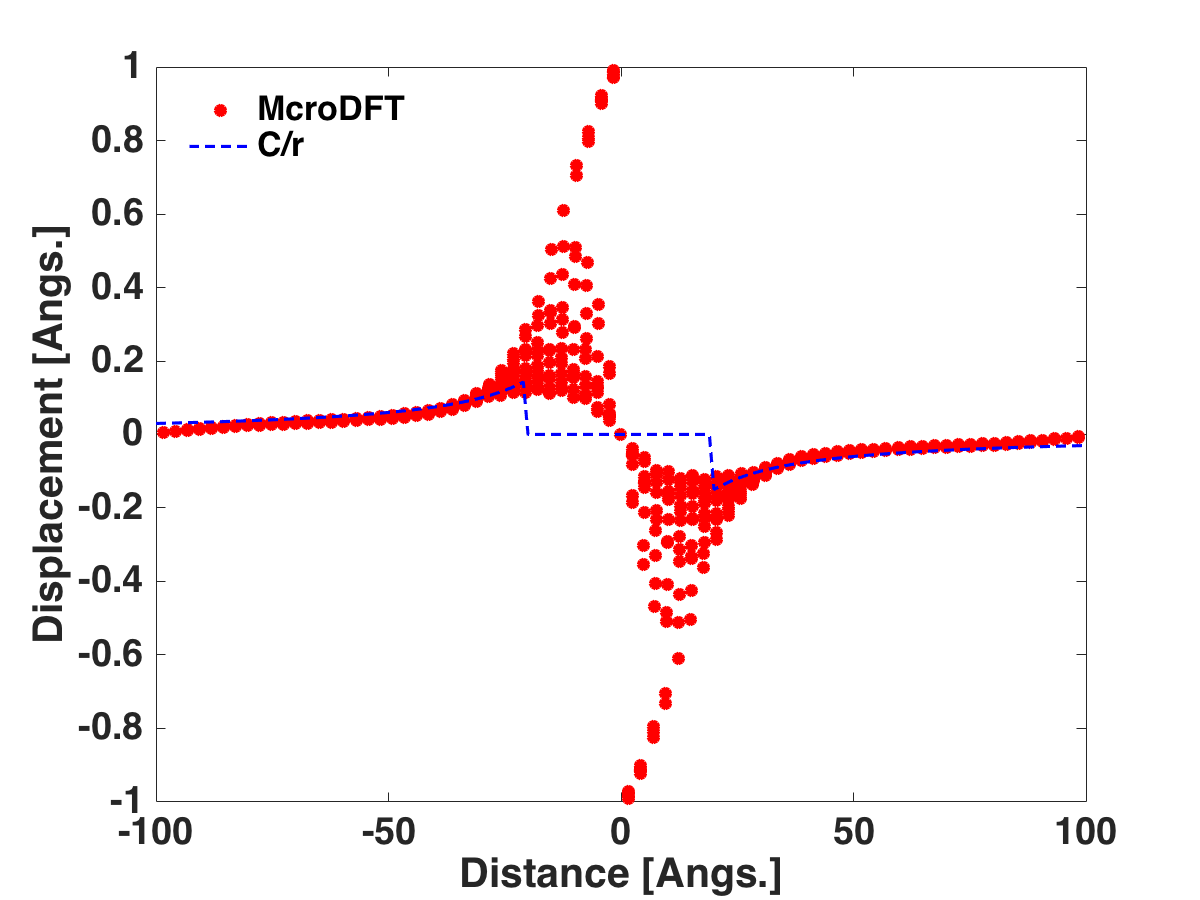}	
	\caption{Atomic displacement with respect to their perfect pristine position near the core for the collapsed PDL containing 37 vacancies. The blue dashed lines indicate a decay of $C/r$ with $C = -3~\AA$. We observe that near the defect, the displacements do not follow the smooth decay. \label{pdl-displacement}}
\end{figure}

\subsection{Screw dislocations in magnesium} 
After relaxing the dislocation core, the excess of energy for the dislocations was measured  with respect to the pristine Mg configuration.  To measure the energy of the dislocation core, we took a cylinder of radius $R$ centered in the dislocation core ($y = 0$, $z =0$) with principal axis along the dislocation line. Then, we analyzed the excess of energy for different values of $R$. The excess energy was defined as

\begin{equation} \label{excess}
E_{\text {excess}}(N,R) = E_{\text{D}}(N,R) - E_{\text{P}}(N,R) = E_{\text{D}}(N,R) - N E_{\text{C}},
\end{equation}
where $N$ is the total number of atoms in the cylinder of radius $R$, $E_{\text{D}}(N,R)$ is the total energy in the same cylinder after the core has been relaxed, and $E_{\text{C}}$ is the cohesive energy of pristine Mg. The energy in the cylinder, $E_{\text{D}}(N,R)$ is computed by evaluating the electronic fields for each ESP inside the cylinder using the predictor-corrector approach. Once the fields are evaluated, the total energy was computed using Eq. \ref{eq:free_cg_sampling}.

\begin{figure} 
\centering
\subfigure[Excess of energy for a basal screw dislocation.]{\includegraphics[width=0.45\textwidth]{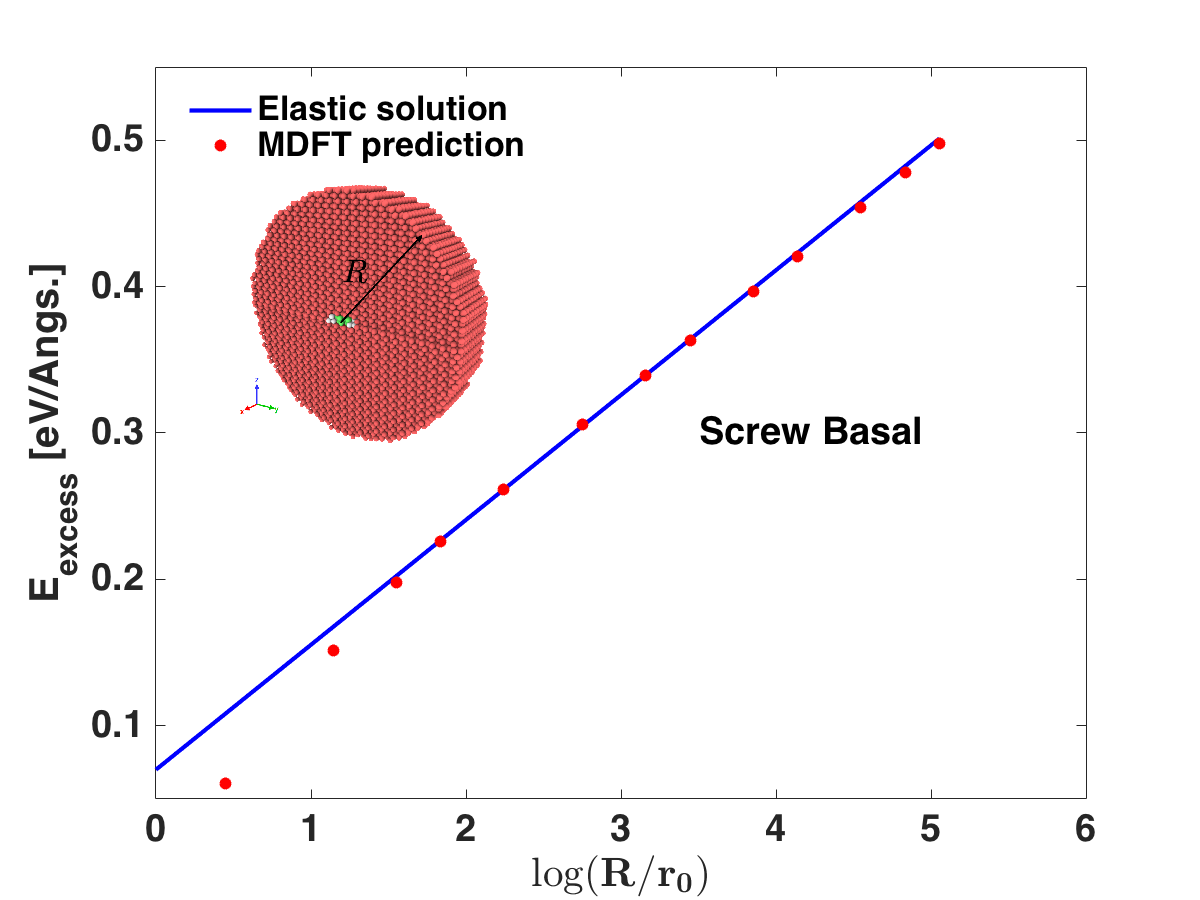}}
\subfigure[Excess of energy for a prismatic screw dislocation.]{\includegraphics[width=0.45\textwidth]{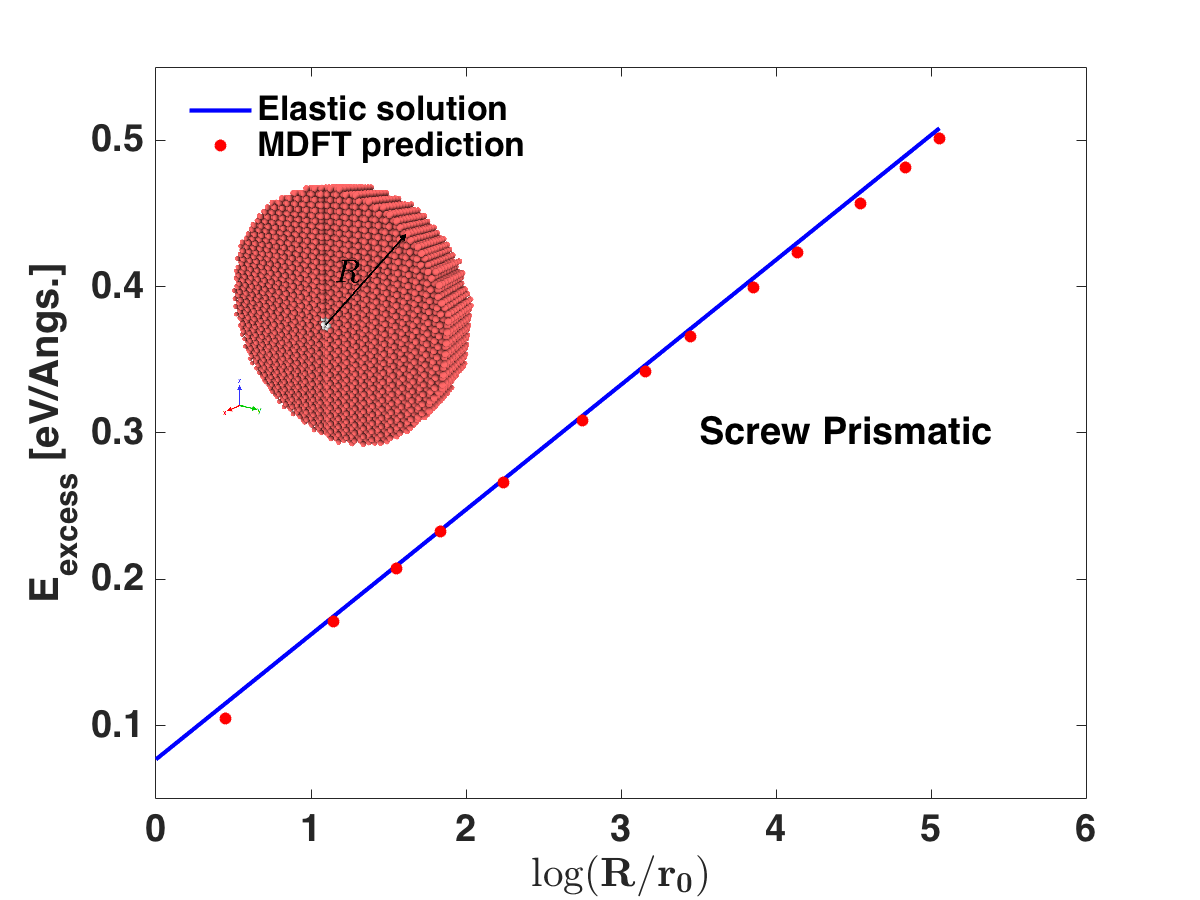}}
\caption{Excess of energy per Angstrom for a) $\dfrac{a_0}{3}[11\overline{2}0]\lbrace0001\rbrace$ and b) $\dfrac{a_0}{3}[11\overline{2}0] \lbrace10\overline{1}0 \rbrace$ screw dislocation. \label{Fig:2} }
\end{figure}

In the analysis of dislocations it is usually common to show the evolution of the excess energy in the cylinder as a function of the ratio $\frac{R}{r_0}$, where $r_0$ is an arbitrary constant. It is common practice to take $r_0=b$, and under this assumption the energy contained in the cylinder can be expressed as \cite{Hirth}


\begin{equation} \label{elastic-solution}
\frac{W}{L} = \frac{\mu b^2}{4 \pi} \ln \left( \frac{R}{r_0} \right) + E_{\text{core}},
\end{equation}
    where $\mu = 19.9$ GPa is the shear modulus  of Mg computed with MacroDFT, and $ E_{\text{core}}$ is an indeterminate constant that needs to be evaluated using discrete techniques such as \textit{ab-initio} or molecular dynamics calculations. Remarkably, the value $E_{\text{core}}$ is not a physical quantity as it depends on the value $r_0$ that is arbitrary. The logarithmic divergence of Eq. \ref{elastic-solution} indicates that in order to understand the energy of the dislocation core, the energy has to be sampled across very large samples spanning multiple Burgers vectors in order to obtain the \emph{elastic} pre-factor $\frac{\mu b^2}{4 \pi}$. 
    
Figure \ref{Fig:2} shows the evolution of the excess energy, computed with Eq. \ref{excess}, as a function of the $\log(R/r_0)$. By adjusting the $E_{\text{core}}$ values to the results obtained with MacroDFT, we have concluded that $E_{\text{core}} = 7$ meV/$\AA$ and $E_{\text{core}} = 7.5$ meV/$\AA$ for the basal and prismatic slip systems, respectively. This small strain energy per unit length is very difficult to measure and illustrate the importance of coarse-grained techniques such as MacroDFT. The results also show an excellent agreement for the far field with the elastic solution (characterized by the slope of the blue curve) which lends credence to the prediction of dislocation far field obtained with MacroDFT and therefore, to the coarse-grained scheme used.

\begin{figure}
\centering
\subfigure[Basal screw dislocation.]{\includegraphics[width=0.45\textwidth]{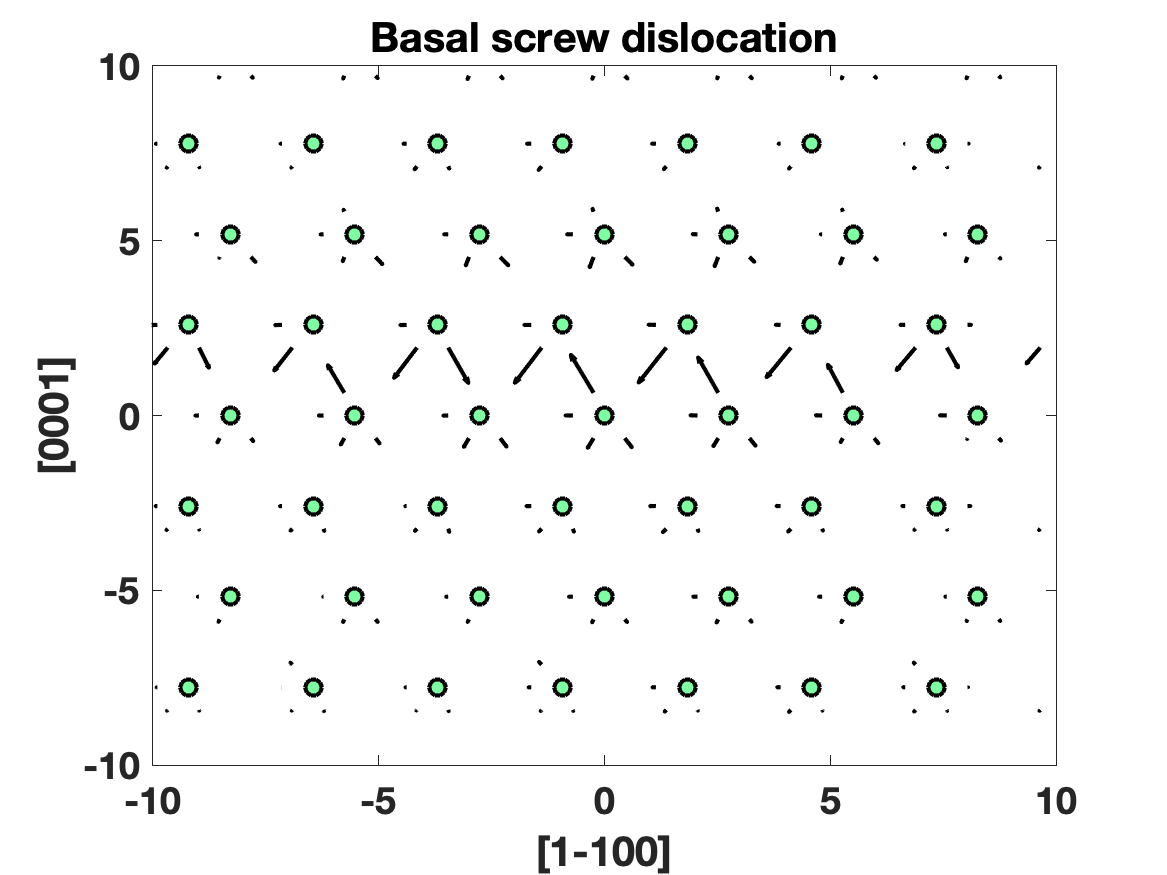}}
\subfigure[Prismatic screw dislocation.]{\includegraphics[width=0.45\textwidth]{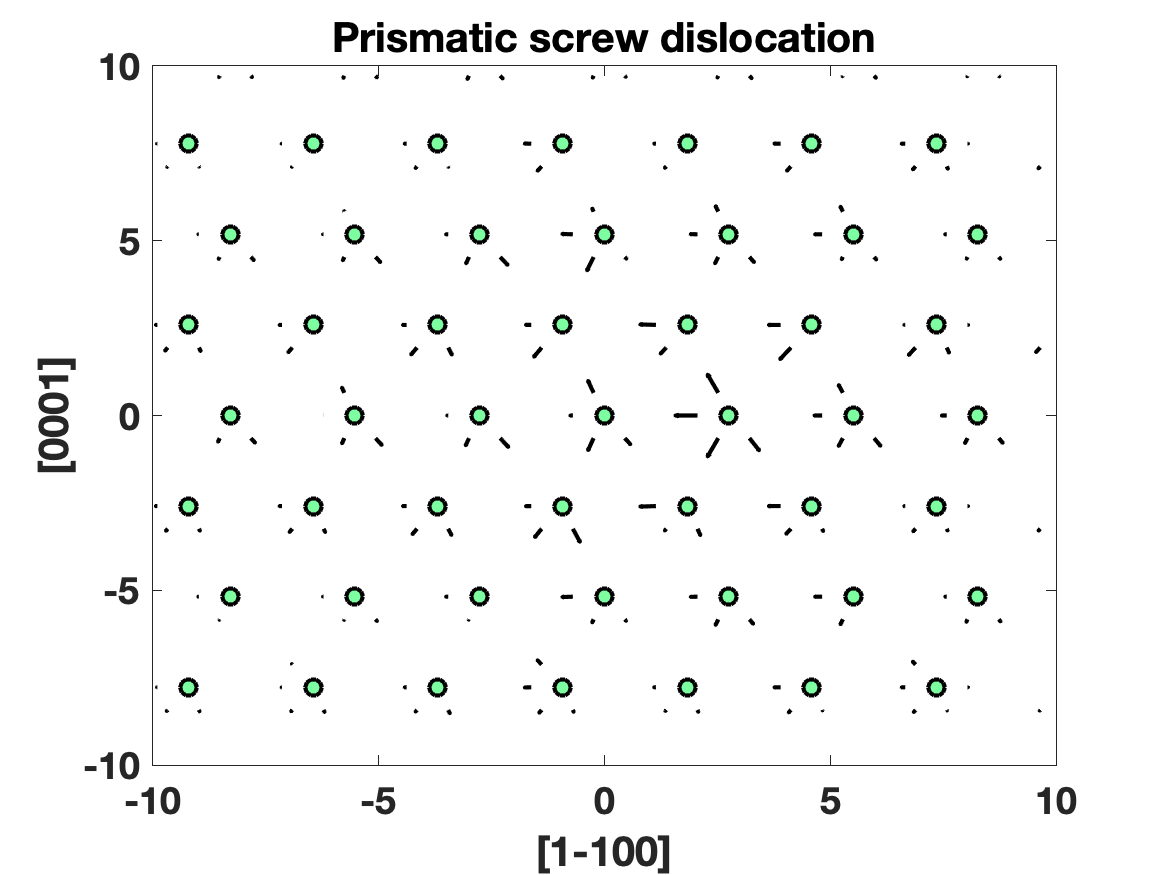}}
\caption{Differential displacement plots for the a) basal and b) prismatic screw dislocation in Mg. The atoms are shown in green, while the arrows indicate the magnitude of the differential displacement field. The prismatic dislocation has a very compact core, while the core in the basal plane is slighter larger due to the existence of stable stacking fault in Mg. \label{Fig:DDplot} }
\end{figure}

Figure \ref{Fig:DDplot} shows the differential displacement plots \cite{Vitek:1970} structure of the dissociated core for the basal $\dfrac{a_0}{3}[11\overline{2}0]\lbrace0001\rbrace$ and prismatic $\dfrac{a_0}{3}[11\overline{2}0] \{ 10\overline{1}0\}$ screw dislocations. The original basal dislocation is dissociated into two partials according to the reaction
\begin{equation} \label{dissociation}
\frac{a_0}{3}[11\overline{2}0] \longrightarrow \frac{a_0}{3}[10\overline{1}0] + \frac{a_0}{3}[01\overline{1}0] + \text{SF},
\end{equation}
where SF stands for stacking fault. By examining the atomic positions as shown in Figure \ref{Fig:DDplot}(a) we observed that the dissociation distance was around $\sim 2.9 a_0$. By comparing with previous studies of screw dislocations in the basal plane, we found that this value is slightly larger than previously reported \cite{Yasi_2009}. This could be attributed to a number of reasons, including different boundary conditions, pseudo-potential used, etc. The $\dfrac{a_0}{3}[11\overline{2}0]\lbrace10\overline{1}0\rbrace$ screw dislocation is stable and \emph{does not split} into two partials. This is consistent with stacking fault calculations along the prismatic plane. These calculations show that there is not stable local minima in this direction, and therefore, the prismatic dislocation cannot split. 

\subsection{Aluminum solute-dislocation interaction}
We now proceed to measure the interaction energy between an Al solute atom and the dislocations for various position along the slip plane. In order to do so, we replace a Mg atom by an Al solute stating from the position $y=0,~z=0$. We only replace atoms lying on the slip plane for the two screw dislocations considered in this work. Then, the excess energy was computed as

\begin{equation}
\Delta E_{\text{solute}}= E_D(N,R,r^{\text{Al}}) -E_D(N,R,r^{\text{Al}}_{\text{max}}) 
\end{equation}
where $E_D(N,R,r^{\text{Al}})$ is the energy in a cylinder of radius $R = 500 \AA$, containing N-1 Mg atoms and one Al atom at $r^{\text{Al}}$, and $E_D(N,R,r^{\text{Al}}_{\text{max}})$ is the energy in a cylinder of radius $R = 500 \AA$, containing N-1 Mg atoms and one Al atom at the $r^{\text{Al}}_{\text{max}}$, which is the maximum distance from the core at which we have placed the solute atom. 

Figure \ref{SoluteExcessEnergy} shows the excess of energy for various positions of the solute atom for the basal and prismatic slip system (Figure \ref{SoluteExcessEnergy}(a) and (b), respectively). Figure \ref{SoluteExcessEnergy} depicts a non-linear interaction energy between solute and dislocation core, and this interaction is very strong near the core decaying very fast relatively quickly. Remarkably, the maximum interaction energy is of the order of 50-60 meV/Angs. which is larger than the core energy of the dislocations. 

\begin{figure} 
\centering
\subfigure[Basal screw.]{\includegraphics[width=0.45\textwidth]{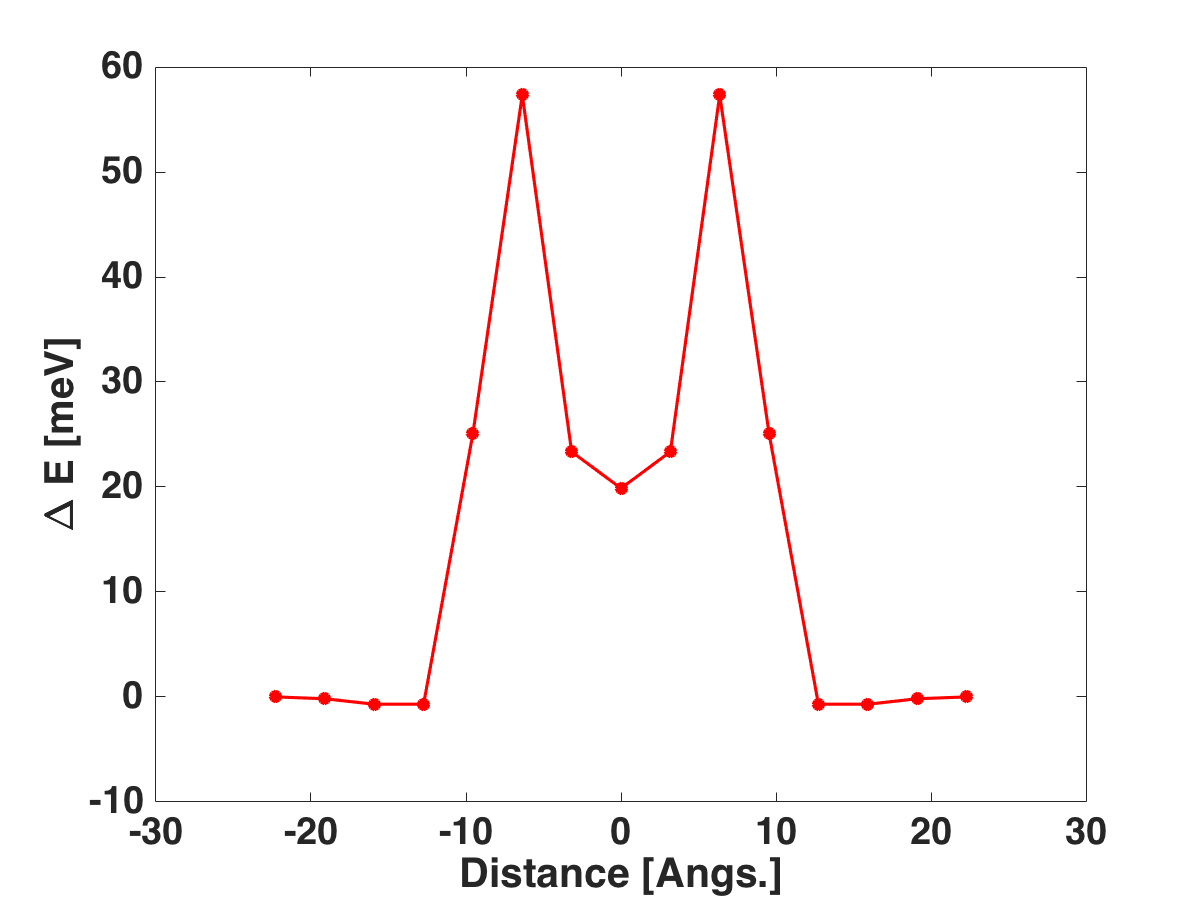}} 
\subfigure[Prismatic screw.]{\includegraphics[width=0.45\textwidth]{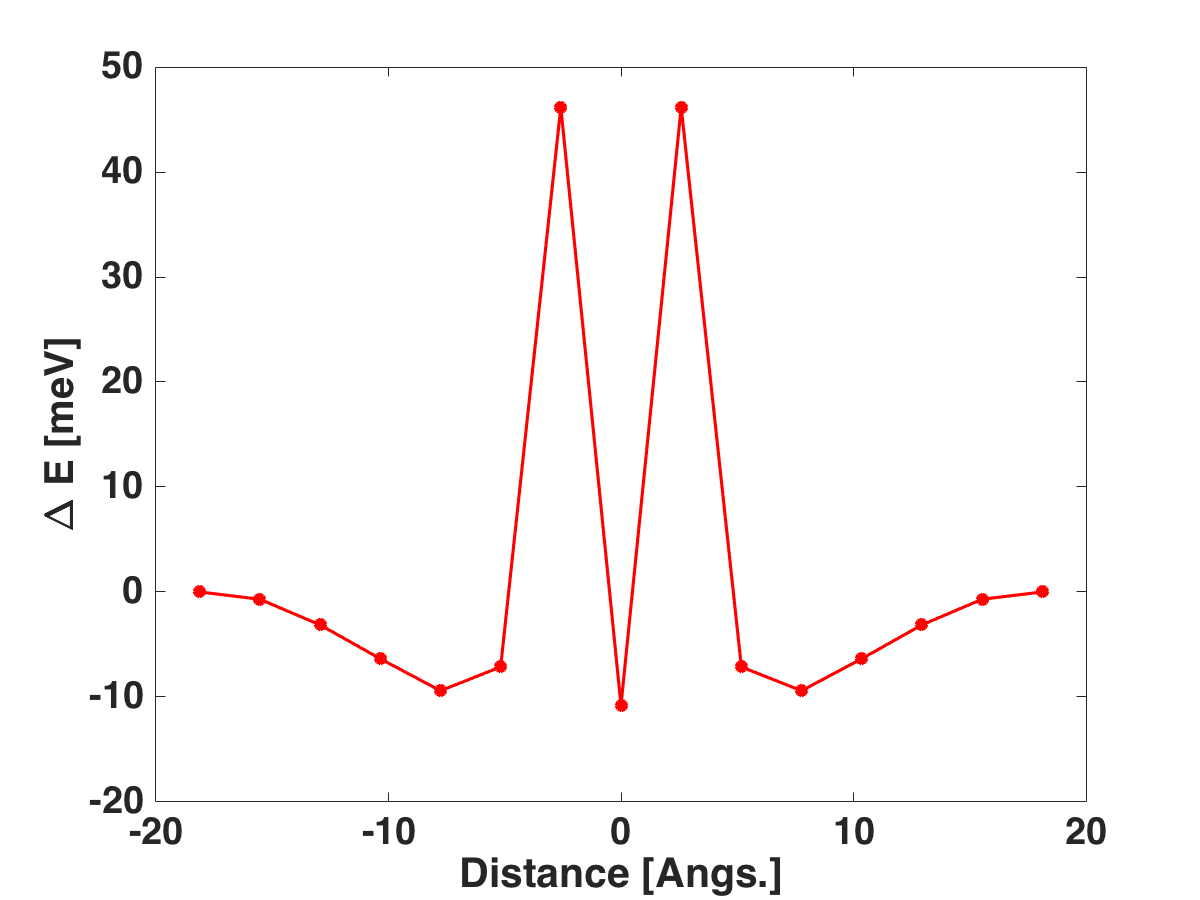}}
\caption{ Excess of energy for an Al solute near the dissociated core along the (a) basal for a screw dislocation lying on the basal plane and (b) prismatic plane for a screw dislocation lying on the prismatic plane. \label{SoluteExcessEnergy}}
\end{figure}


\section{Parallel performance} \label{parallel}

\begin{figure} 
\centering
\includegraphics[width=0.75\textwidth]{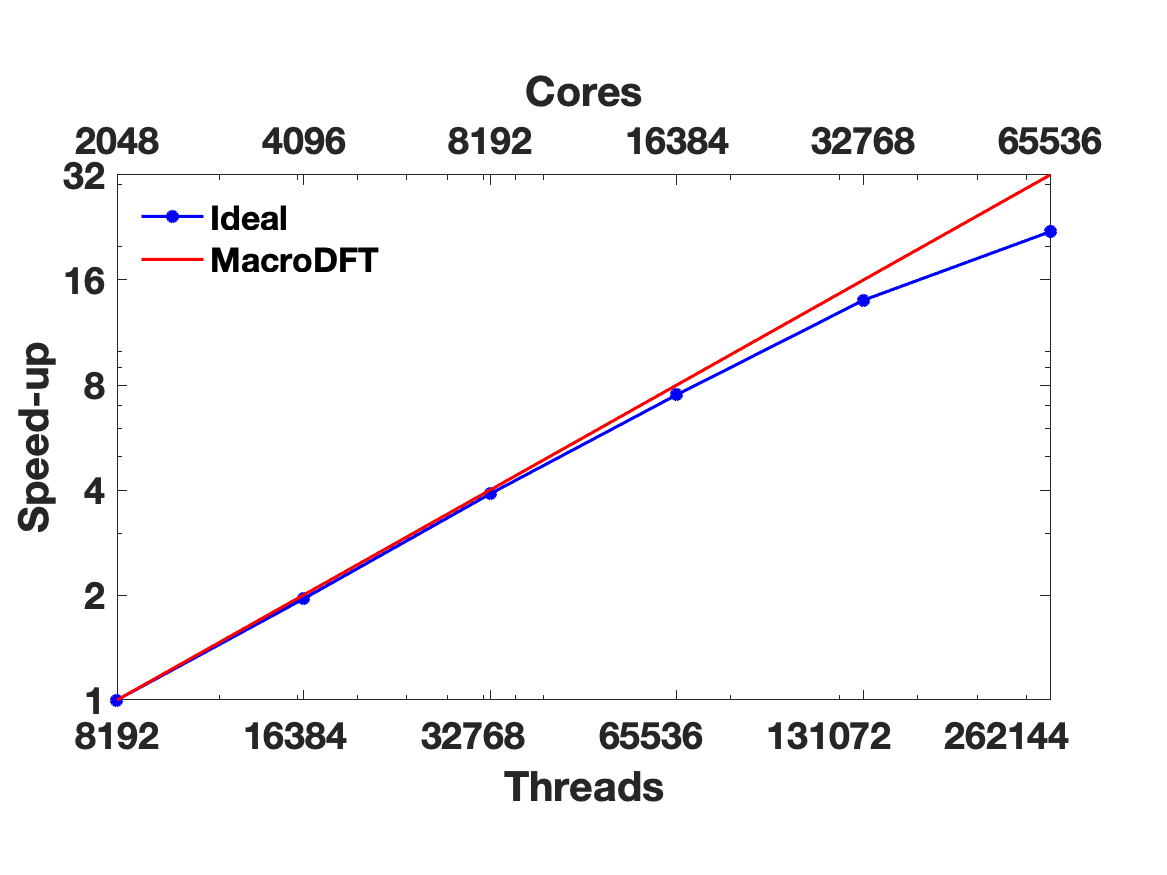}
\caption{Speed-up achieved with multiple combinations of processors and threads obtained in MIRA IBM BG/Q 1.6 GHz PowerPC A2 supercomputer of Argonne National Laboratory. We use a combination of 64 threads per node, thus we use 2,048 to 65,536 processors (128 to 4096 nodes). Each node has 16 physical processors and capable of running 64 threads per node. \label{Scaling}}
\end{figure}

We now proceed to discuss the parallel performance and speed-up achieved with MacroDFT in MIRA an IBM BG/Q 1.6 GHz PowerPC A2 supercomputer property of Argonne National Laboratory. Measuring the efficiency and performance are essential for new DFT methodologies and codes, as most of the classical plane-wave codes scale poorly for more than 64 processors. Thus, the \emph{next generation} of computational tools should be able to harness petascale resources that can further accelerate the discovery of new materials and alloys through accurate \emph{ab-initio} calculations. This is challenging since both the methodology and implementation need to be highly efficient to handle thousands of processors at the same time. Here, we provide results for the strong scaling of MacroDFT using up to 262,144 threads. The IBM BG/Q 1.6 GHz PowerPC A2 node architecture has 16 physical processors, each node is capable of running 4 threads. Thus, the user can specify any combination of threads and processor per node such that the number $\text{\# processors }\times\text{ \# threads} = 64$. The ratio processors/threads has to be evaluated for each code/problem. In our case, we found that the best combination was to use processes with 64 threads, which reduces the internode communication. Using this combination, the threads share the same memory, the number of MPI communications was reduced significantly. For the benchmark, we have used the non-collapsed cluster of vacancies containing 7 vacancies described in Section \ref{PDL-section} and compute one SCF evaluation using different numbers of threads. 

Figure \ref{Scaling} shows the parallel performance for MacroDFT from 8,192 (2,048) to 262,144 (65,536) threads (processors). The results show that MacroDFT has a good efficiency up to 131,072 threads with an efficiency of $\sim87.5\%$, while for  262,144 threads, an efficiency of $\sim 68.75\%$ is achieved. This remarkable performance is due to the fact that the DFT formulation (Eq. \ref{eq:9}) can be solved \emph{locally} and the relevant electronic fields ESP can be evaluated with only the information of a small neighborhood surrounding the point. Since the computational representation of the ESPs is in real-space, the cost of computing the electronic fields can be distributed between multiple processors taking advantage of massively parallel computers, as only the local components of the Hamiltonian need to be updated every SCF iteration. According to the scaling performance shown in Figure \ref{Scaling}, the time taken by the interpolation scheme is negligible in comparison with the time consumed to determine the nodes and weights of the spectral quadrature rule. By contrast, when the number of threads is increased significantly, 262,144 or more, the performance goes down due to multiple factors including load imbalance, few number of ESP per node, interpolation, and data writing to disk. Nonetheless, the performance of MacroDFT is outstanding when compared with the \emph{state-of-the-art} DFT codes, which cannot even run in such a large number of processors. Therefore, real-space DFT implementations have an opportunity to take advantage of supercomputing power and over-perform traditional plane-wave based implementations.

\section{Conclusions} \label{Conclusions}
We have developed a novel approach for the study of dislocations in crystalline materials using \emph{ab-initio} techniques.  In this approach, the equations of density functional theory are solved on a domain large enough to capture both the core and elastic fields in a completely seamless manner.  It technique is based on three-main pillars i) reformulation of the Kohn-Sham equations into a density matrix formulation and subsequent spectral representation of the density matrix (See Section \ref{sec:dft}); ii) Separation of the electronic fields in predictor and corrector counterparts (Equations (\ref{eq:pc}), (\ref{phi_approx_scheme}) and Section \ref{sec:pred}); and iii) Interpolation scheme for electronic fields and displacements using finite element meshes (\ref{sec:cg}). The proposed approach has been implemented in the MacroDFT code \cite{Ponga:2016} which can be executed in MPI, thread and hybrid options according to the architecture of the high performance cluster available. The parallel efficiency of the code has been studied, achieving good parallel performance in extreme scale computers capable of running $\sim 260,000$ threads. 

The code and method has been used to study the relaxed configurations of the atoms in multiple dislocation cores in Mg, a crystalline metallic material. For PDLs, our simulations have predicted that vacancies always bind together, at least for low temperatures where entropic effects are weak. In screw dislocations gliding in the basal and prismatic planes, we have found that the strain energy stored in the computational sample scales accordingly to the elasticity theory for both cases. The results indicate that in order to obtain good prefactors, the simulations has to be sufficiently large, of the order of a few hundred thousand atoms. The agreement of the elastic prefactor between our simulations and theory lends credence to the approach, and provides numeric evidence that the coarse-grained approach can be used in complex scenarios where quantum mechanical interactions are important, such as in dislocation cores in alloys, grain and twin boundaries. 
%

\section{Acknowledgments}
We gratefully acknowledge the support of the U.S. Army Research Laboratory (ARL) through the Materials in  Extreme Dynamic Environments (MEDE) Collaborative Research Alliance (CRA) under Award Number W911NF-11-R-0001 and from the Natural Sciences and Engineering Research Council of Canada (NSERC) through the Discovery Grant under Award Application Number RGPIN-2016-06114. We also are grateful to Compute Canada through the Westgrid consortium for giving access to the supercomputer grid. This research used resources of the Argonne Leadership Computing Facility, which is a DOE Office of Science User Facility supported under Contract DE-AC02-06CH11357.

\bibliographystyle{unsrt}
\bibliography{macrodft-dislocations}

\end{document}